\documentclass{article}
\usepackage[utf8]{inputenc}
\usepackage{amssymb}
\usepackage{amsmath}
\usepackage{amsfonts}
\usepackage[onehalfspacing]{setspace}
\usepackage{geometry}%
\usepackage{graphicx}
\providecommand{\U}[1]{\protect\rule{.1in}{.1in}}

\begin{document}

\title{Error Rate Analysis of Amplitude-Coherent Detection over Rician Fading
Channels with Receiver Diversity}
\author{Mohammad Al-Jarrah$^{\dag}$, Ki-Hong Park$^{\ddag}$, Arafat Al-Dweik$^{\dag}$,
and Mohamed-Slim Alouini$^{\ddag}$\\$^{\dag}$Khalifa University of Science and Technology, Abu Dhabi, UAE. \\E-mail: \{mohammad.aljarrah, arafat.dweik\}@ku.ac.ae.\\$^{\ddag}$King Abdullah University of Science and Technology (KAUST), Thuwal,
Kingdom of Saudi Arabia. \\E-mail: \{ kihong.park, slim.alouini\}@kaust.edu.sa.}
\maketitle

\begin{abstract}
Amplitude-coherent (AC) detection is an efficient detection technique that can
simplify the receiver design while providing reliable symbol error rate (SER).
Therefore, this work considers AC detector design and SER analysis using
$M$-ary amplitude shift keying (MASK) modulation over Rician fading channels.
More specifically, we derive the optimum, near-optimum and a suboptimum AC
detectors and compare their SER to the coherent, noncoherent and the heuristic
AC detectors. Moreover, the analytical SER of the heuristic detector is
derived using two different approaches for single and multiple receiving
antennas. One of the derived expressions is expressed in terms of a single
integral that can be evaluated numerically, while the second approach gives a
closed-form analytical expression for the SER, which is also used to derive a
simple formula for the asymptotic SER at high signal-to-noise ratios (SNRs).
The obtained analytical and simulation results show that the SER of the AC and
coherent MASK detectors are comparable, particularly for high values of the
Rician $K$-factor, and small number of receiving antennas. Moreover, the
obtained results show that the SER of the optimal AC detector is equivalent to
that of the coherent detector. However, the optimal AC\ detector complexity is
prohibitively high, particularly at high SNRs. In most of the scenarios, the
heuristic AC detector significantly outperforms the optimum noncoherent
detector, except for the binary ASK case at low SNRs. Moreover, the obtained
results show that the heuristic AC detector is immune to phase noise, and
thus, it outperforms the coherent detector in scenarios where system is
subject to considerable phase noise.

\end{abstract}

keywords: Optical wireless communications (OWC), free space optics (FSO), non-coherent,
semi-coherent, amplitude-coherent, Rician, Ricean, receiver diversity, phase noise.

\section{Introduction}

\markboth{Error Rate Analysis of Amplitude-Coherent Detection}
{Murray and Balemi: Using the Document Class IEEEtran.cls}%
Generally speaking, there are three main types of detection schemes for
digital signals, which are coherent detection, noncoherent detection, and
partially coherent detection \cite{Dweik}. The detector design, required
channel state information (CSI), computational complexity, and symbol error
rate (SER) of each detection scheme depend on several factors such as the
modulation scheme, modulation order, and channel model. Therefore, adopting a
particular modulation and detection schemes is mostly determined by the
targeted application. For example, broadband communications require spectrally
efficient modulation schemes to support high data rates, and the communicating
nodes typically have sufficient resources to estimate the CSI, and hence,
modulation schemes with high order and coherent detectors are utilized. For
most wireless applications, quadrature amplitude modulation (QAM) is
considered as the most attractive due to its power and spectral efficiency
\cite{DVB}-\cite{LTE-A}. Nevertheless, $M$-ary amplitude shift keying (MASK)
has recently attracted extensive attention because it is more suitable for
certain applications such as wireless sensor networks (WSNs) \cite{WSN},
wireless energy transfer \cite{NCD1}, radio frequency identification (RFID)
\cite{RFID}, and optical wireless communications (OWC) \cite{FSO4}-\cite{FSO3}.

Unlike typical wireless communications systems, OWC such as free space optics
(FSO) and visible light communications (VLC) that use intensity modulation
with direct detection (IM-DD) require the baseband signal to be real and
positive, and hence, using QAM for OWC directly is infeasible. To overcome
this limitation, QAM can be combined with orthogonal frequency division
multiplexing (OFDM) to generate real and positive signals using various
techniques \cite{Uysal-Survey}. Nevertheless, the spectral efficiency and SER
of QAM-OFDM is generally equivalent to MASK-OFDM \cite{Haas-Sep-2013}.
Therefore, MASK renders itself as an efficient alternative to QAM for OWC
\cite{Lu-4PAM} because it can be used with/without OFDM. Moreover, in IM-DD,
the binary ASK (BASK) can be detected using a simple noncoherent detector that
does not require prior knowledge of the instantaneous CSI. The noncoherent
BASK detector has low complexity and robust to hardware impairments such as
the carrier frequency offset and phase noise, but it suffers from poor
spectral efficiency, and accurate knowledge of the statistical CSI is
necessary to compute the optimum threshold. Improving the spectral efficiency
of noncoherent MASK modulation by increasing the modulation order $M$ is not
feasible in fading channels due to its poor symbol error rate (SER)
\cite{Dweik}, which limits its utilization to additive white Gaussian noise
(AWGN) channels. Practically speaking, the AWGN channel model is limited to
few applications such as indoor OWC \cite{Uysal-Survey}.

Although the channel in OWC may be considered non-fading in certain scenarios
\cite{Armstrong-2008}, \cite{Zhang-2014}, channel models that consider the
fading induced by atmospheric turbulence can be considered more practical, and
they are actually more flexible because they can be used to describe a wide
range of fading scenarios. In the literature, several channel models have been
adopted for OWC including the Gamma-Gamma, exponential and Rician
\cite{Uysal-Survey}, \cite{Letzepis-2009}, \cite{Aghajanzadeh-2010}. Moreover,
the random pointing error in OWC are typically modeled as Rician
\cite{Slim-ICC-2018}-\cite{Ansari-2014}. The Rician channel model is of
particular interest because it is also widely adopted in wireless
communications systems, such as massive multiple-input multiple-output (MIMO)
systems \cite{Yue-2015}-\cite{Sun-2016}, and satellite/drone to ground
channels \cite{Matolak-II}, \cite{Matolak-III}. Therefore, the Rician channel
model is adopted in this work.

To resolve the spectral efficiency and SER conflict of MASK, Al-Dweik and
Iraqi \cite{Dweik} recently proposed a semi-coherent detection scheme, also
denoted as amplitude coherent (AC) detection, that allows using MASK
modulation with $M>2$ over dispersive channels while maintaining the main
advantages of noncoherent detection such as receiver low complexity, and
immunity to phase noise and frequency offsets. The AC detector requires only
the knowledge of the channel gain, which can be obtained blindly and
efficiently for single and multicarrier modulation schemes \cite{Bariah-TVT}.
The channel phase information is not required, which is the main factor that
contributes to the complexity reduction of the detector. The optimum,
suboptimum and a heuristic detectors are derived in Rayleigh fading channels,
and the performance of the heuristic detector is evaluated with and without
perfect knowledge of the channel gain in \cite{Dweik} and \cite{Bariah-TVT},
respectively. The optimum amplitudes of the transmitted MASK symbols are then
derived for multibranch detectors in \cite{Malik}. However, the Rayleigh
fading model is limited to wireless applications with no line-of-sight (LoS)
signal component. Therefore, applying the AC detector and evaluating its
performance in a more general channel model is indispensable.

Consequently, this paper considers applying the AC detection technique to
communications system in Rician fading channels with single and multiple
receiving antennas. More specifically, the optimum AC detector is derived, and
its SER is compared with the optimum coherent and noncoherent MASK detectors.
Moreover, the SER of the heuristic detector \cite{Dweik} is derived using two
different approaches, and efficient expressions are obtained. One of the
approaches results in closed-form SER formula, which is then simplified to
provide the asymptotic SER at high SNRs. The other approach results in an
efficient expression that contains a single integral. The obtained analytical
and simulation results demonstrate that the AC detector can offer reliable SER
performance that is comparable to coherent detection in Rician fading channels.

The rest of the paper is organized as follows. Section II presents the model
of $M$-ary ASK system. In section III, the different types of considered
detectors are derived including coherent, non-coherent and amplitude coherent
detectors. Sections IV and V present the two approaches considered for
analyzing the SER for the heuristic AC detector. Sections VI and VII provide
the numerical results and conclusion, respectively.

\section{System and Channel Models}

In unipolar MASK systems, the baseband representation of the transmitted
signal during the $\ell$th signaling interval is given by
\begin{equation}
d^{\{\ell\}}=s_{m}\text{, }m\in\{0\text{, }1\text{, }...\text{, }M-1\},
\end{equation}
where $M$ is the modulation order, the transmitted symbols $s_{m}\in
\mathbb{R}$, where $\mathbb{R}$\ the set of positive real numbers including
the $0$. Without loss of generality, the symbols' amplitudes can be ordered
such that $s_{m+1}>s_{m}$. Moreover, the amplitude spacing is assumed to be
uniform such that $s_{m+1}-s_{m}=\delta$. It should be noticed that $\frac
{1}{M}\sum_{m=0}^{M-1}E_{m}=1$, $E_{m}=s_{m}^{2}$ when the average symbol
energy is normalized to unity. Therefore, the transmitted symbol during the
$\ell$th transmission interval can be described by,%
\begin{equation}
d^{\left\{  \ell\right\}  }=m\times\delta\text{, \ \ }m\in\{0\text{, }1\text{,
}...\text{, }M-1\},
\end{equation}
where $m$ is selected uniformly, and%
\begin{equation}
\delta=\sqrt{\frac{6}{\left(  2M-1\right)  \left(  M-1\right)  }}\text{.}%
\end{equation}

The system under consideration assumes that the transmitter is equipped with
single transmit antenna, and the receiver is equipped with $N$ receiving
antennas. The channels between the transmitting and receiving antennas are
assumed to be flat, independent and identically distributed (iid) Rician
fading channels. Therefore, the received signals in vector notations can be
written as%
\begin{equation}
\mathbf{r}=\mathbf{h}s_{m}+\mathbf{n},\label{E-r-1}%
\end{equation}
where the channel fading vector $\mathbf{h}\in\mathbb{C}^{N\times1}$,
$h_{i}\sim\mathcal{CN}\left(  m_{h}\text{, }2\sigma_{h}^{2}\right)  $
represents the Rician fading, $s_{m}$ is the information symbol selected
uniformly from the set $\mathbb{S=}\left\{  s_{0}\text{, }s_{1}\text{, ...,
}s_{M-1}\right\}  $, and the additive white Gaussian noise (AWGN)$\ $vector
$\mathbf{n}\in\mathbb{C}^{N\times1}$ where $n_{i}\sim\mathcal{CN}\left(
0\text{, }2\sigma_{n}^{2}\right)  $. The received signal in (\ref{E-r-1}) can
also be written as
\begin{equation}
\mathbf{r}=\left[  \mathbf{\alpha}\circ\mathbf{\Phi}\right]  s_{m}+\mathbf{n},
\end{equation}
where $\mathbf{\alpha}=\left[  \left\vert h_{1}\right\vert \text{, }\left\vert
h_{2}\right\vert \text{, }\ldots\text{, }\left\vert h_{N}\right\vert \right]
$, $\mathbf{\Phi}\triangleq e^{j\mathbf{\theta}}$, and $\circ$ denotes the
Hadamard product.

\subsection{Rician Channel Model}

In Rician fading, the received signal has a LoS component that affects the
received signal envelope and phase. After dropping the channel index, the
joint probability density function (PDF) of the channel envelope
$\alpha\triangleq\left\vert h\right\vert $ and and phase $\theta\triangleq
\arg\left\{  h\right\}  $ is given by%
\begin{equation}
f\left(  \alpha,\theta\right)  =\frac{\alpha}{2\pi\sigma_{h}^{2}}\exp\left(
-\frac{\alpha^{2}-2\mu_{h}\alpha\cos\theta+\mu_{h}^{2}}{2\sigma_{h}^{2}%
}\right)  ,\label{joint}%
\end{equation}
where $\mu_{h}=\left\vert m_{h}\right\vert $. The marginal PDF of $\alpha$ can
be obtained by averaging the joint PDF $f\left(  \alpha,\theta\right)  $ over
$\theta$. Thus%
\begin{align}
f\left(  \alpha\right)   & =\int_{-\pi}^{\pi}f\left(  \alpha,\theta\right)
\text{ }d\theta\label{env}\\
& =\frac{2(1+K)}{\Omega}\alpha\operatorname{e}^{-K}\operatorname{e}%
^{-\frac{(1+K)}{\Omega}\alpha^{2}}\text{ }I_{0}\left(  2\alpha\sqrt
{\frac{K(1+K)}{\Omega}}\right)  ,
\end{align}
where $\Omega=\mu_{h}^{2}+2\sigma_{h}^{2}$ and $K=\frac{\mu_{h}^{2}}%
{2\sigma_{h}^{2}}$. Similarly, the PDF of the phase $\theta$ can be obtained
by averaging over the PDF of $\alpha$ \cite{Rice-phase}, which gives%
\begin{equation}
f(\theta)=\frac{1}{2\pi}\exp\left(  -\frac{\mu_{h}^{2}}{2\sigma_{h}^{2}%
}\right)  +\frac{\mu_{h}\cos\left(  \theta+\phi\right)  }{\sqrt{2\pi}%
\sigma_{h}}\exp\left(  \frac{-\mu_{h}^{2}}{2\sigma_{h}^{2}}\sin^{2}\left(
\theta+\phi\right)  \right)  Q\left(  -\frac{\mu_{h}}{\sigma_{h}}\cos\left(
\theta+\phi\right)  \right)  ,\label{phase}%
\end{equation}
where $\phi=\tan^{-1}\left(  \frac{\mu_{h,Q}}{\mu_{h,I}}\right)  $, $\mu
_{h,I}\triangleq\Re\left\{  m_{h}\right\}  $ and $\mu_{h,Q}=\Im\left\{
m_{h}\right\}  $.

\section{MASK Detector Design}

Usually, there is a trade-off between the receiver complexity and SER
performance. The complexity may refer to the computational complexity,
hardware complexity or the amount of information required at the receiver
side. Adopting a certain detector design depends on the desired application.
Other parameters such as the spectral efficiency can affect the complexity and
SER. Because it is typically difficult to achieve such conflicting objectives
simultaneously, it is crucial to have various options that may fit various
applications. In this section, various optimum and suboptimum detectors are
derived for MASK signals in Rician fading channels, and their complexity will
be discussed.

\subsection{Coherent Detection}

Based on the signal model in (\ref{E-r-1}), and noting that all received $N$
signals are mutually independent, the conditional PDF of $\mathbf{r}$ for a
given fading vector $\mathbf{h}$, and a transmitted symbol $s_{m}$ is given by
\cite{alouni}%
\begin{equation}
f\left(  \mathbf{r|h,}s_{m}\right)  =\prod\limits_{i=1}^{N}f(r_{i}%
\mathbf{|}h_{i}\text{,}s_{m}).
\end{equation}
As can be noted from (\ref{E-r-1}), $f(r_{i}\mathbf{|}h_{i}$,$s_{m}%
)\sim\mathcal{CN}\left(  h_{i}s_{m},2\sigma_{n}^{2}\right)  $, and thus
\begin{align}
f\left(  \mathbf{r|h,}s_{m}\right)   & =\prod\limits_{i=1}^{N}f(r_{i}%
\mathbf{|}h_{i}\text{,}s_{m})\nonumber\\
& =\frac{1}{\left(  2\pi\sigma_{n}^{2}\right)  ^{N}}\prod\limits_{i=1}^{N}%
\exp\left(  -\frac{1}{2\sigma_{n}^{2}}\left\vert r_{i}\mathbf{-}h_{i}%
s_{m}\right\vert ^{2}\right)  .\label{E-PDF-00}%
\end{align}
The maximum likelihood (ML) detector based on (\ref{E-PDF-00}) can be
formulated as%
\begin{equation}
\hat{d}=\arg\max_{\tilde{s}_{m}\in\mathbb{S}}\text{ }f\left(  \mathbf{r|h}%
\text{,}\tilde{s}_{m}\right)
\end{equation}
which gives after taking the $\log$ of the objective function, dropping the
common terms and constants,
\begin{equation}
\hat{d}=\arg\min_{\tilde{s}_{m}\in\mathbb{S}}\sum\limits_{i=1}^{N}\left\vert
r_{i}\mathbf{-}h_{i}\tilde{s}_{m}\right\vert ^{2}\text{.}\label{E-MLD-Coh-00}%
\end{equation}
As can be noted from (\ref{E-MLD-Coh-00}), the computational complexity of the
coherent detector is low, however, the fading parameters represented by
$\mathbf{h}$ should be estimated. Generally speaking, estimating $\mathbf{h}$
requires significant efforts, and inaccurate channel estimation deteriorates
the system SER \cite{Saci}.

\subsection{Noncoherent Detection}

The noncoherent detector can be derived following the same approach of the
coherent detector, except that the detector should not have any information
about the instantaneous values of $\mathbf{h}$. Consequently, $\mathbf{h}$
should be treated as a random vector. In such cases, the ML detector can be
formulated as%
\begin{align}
\hat{d}  & =\arg\max_{\tilde{s}_{m}\in\mathbb{S}}\text{ }f\left(
\mathbf{r|}\tilde{s}_{m}\right) \nonumber\\
& =\arg\max_{\tilde{s}_{m}\in\mathbb{S}}\text{ }\prod\limits_{i=1}^{N}%
f(r_{i}\mathbf{|}\tilde{s}_{m}).
\end{align}
The conditional PDF $f\left(  r_{i}\mathbf{|}s_{m}\right)  $ is the sum of two
complex Gaussian random variables and thus%
\begin{equation}
f\left(  r_{i}\mathbf{|}s_{m}\right)  =\frac{1}{\pi\sigma_{\mathrm{r}}^{2}%
}\exp\left(  -\frac{\left\vert r_{i}\mathbf{-}\mu_{h}s_{m}\right\vert ^{2}%
}{\sigma_{\mathrm{r}}^{2}}\right)  .
\end{equation}
where $\sigma_{\mathrm{r}}^{2}\triangleq2\left(  \sigma_{h}^{2}s_{m}%
^{2}+\sigma_{n}^{2}\right)  $After some straightforward simplifications, the
ML noncoherent detector reduces to%
\begin{equation}
\hat{d}=\arg\min_{\tilde{s}_{m}\in\mathbb{R}}\left\{  -N\ln\left(  \pi
\tilde{\sigma}_{\mathrm{r}}^{2}\right)  +\frac{1}{\tilde{\sigma}_{\mathrm{r}%
}^{2}}\sum\limits_{i=1}^{N}\left\vert r_{i}\mathbf{-}\mu_{h}\tilde{s}%
_{m}\right\vert ^{2}\right\}  ,\label{E-MLD-noncoherent-00}%
\end{equation}
where $\tilde{\sigma}_{\mathrm{r}}^{2}=\sigma_{\mathrm{r}}^{2}$ except that
$s_{m}$ is replaced by $\tilde{s}_{m}$. As can be noted from
(\ref{E-MLD-noncoherent-00}), the noncoherent detector does not require the
knowledge of $\mathbf{h}$, instead, it requires the channel statistical
information, i.e., the values of $\mu_{h}$, $\sigma_{h}^{2}$ and $\sigma
_{n}^{2}$. Estimating the statistical information of the channel is generally
challenging because it requires large number of observations, and hence, large
delay and high computational complexity. Therefore, similar to the coherent
detector, the noncoherent detector has complexity limitations as well.

\subsection{Amplitude Coherent Detection}

The AC detector is designed such as a compromise between the poor SER of the
noncoherent and the high complexity of the coherent detector caused by the
channel estimation process \cite{Dweik}. More specifically, the AC detector is
designed assuming that the receiver has partial knowledge about the channel,
namely, the fading gains vector $\mathbf{\alpha}$, but no information is
required for $\mathbf{\Phi}$. Since phase estimation is typically more complex
to achieve as compared to the channel envelope, the AC detector complexity is
less than the coherent detection \cite{Bariah-TVT}. The following subsections
present the derivation of the optimum and suboptimum AC detectors.

\subsubsection{Optimum AC Detector}

The optimum AC detector can be derived by applying the ML criterion and
assuming the phase shift introduced by the channel is unknown. Therefore,
\begin{equation}
\hat{d}=\arg\max_{\tilde{s}_{m}\in\mathbb{S}}\text{ }f\left(  \mathbf{r|\alpha
,}\tilde{s}_{m}\right)  \text{.}\label{E-ACD-Optimum}%
\end{equation}
Because $r_{i}$ $\forall i$ are mutually independent, then the conditional
joint PDF of $\mathbf{r}$ given that only the channel gain $\mathbf{\alpha}$
is known, can be derived as
\begin{equation}
f\left(  \mathbf{r|\alpha,}s_{m}\right)  =\prod\limits_{i=1}^{N}\int_{-\pi
}^{\pi}f\left(  r_{i}|\alpha_{i}\text{,}\theta_{i}\text{,}s_{m}\right)
f_{\theta_{i}}(\theta_{i})d\theta_{i},\label{E-PDF-01}%
\end{equation}
where $f(\theta_{i})$ is given in (\ref{phase}). By noting that the real and
imaginary parts of $r_{i}$\ are independent, and dropping the index $i$ for
notational simplicity, then $f\left(  r_{i}|\alpha_{i},\theta_{i}%
,s_{m}\right)  $ can be written as%
\begin{align}
f\left(  r|\alpha,\theta,s_{m}\right)   & =f\left(  r_{\Re}|\alpha
,\theta,s_{m}\right)  f\left(  r_{\Im}|\alpha,\theta,s_{m}\right) \nonumber\\
& =\frac{1}{2\pi\sigma_{n}^{2}}\exp\left[  -\frac{\left\vert r\right\vert
^{2}+\alpha^{2}s_{m}^{2}}{2\sigma_{n}^{2}}\right]  \exp\left[  \frac{\alpha
s_{m}}{\sigma_{n}^{2}}\left(  r_{\Im}\sin\left(  \theta\right)  +r_{\Re}%
\cos\left(  \theta\right)  \right)  \right] \nonumber\\
& =\frac{1}{2\pi\sigma_{n}^{2}}\exp\left[  -\frac{\left\vert r\right\vert
^{2}+\alpha^{2}s_{m}^{2}}{2\sigma_{n}^{2}}\right]  \exp\left[  \frac{\alpha
s_{m}}{\sigma_{n}^{2}}\left\vert r\right\vert \cos(\theta-\theta_{\text{r}%
})\right]  ,\label{E-PDF-02}%
\end{align}
where $\theta_{\text{r}}\triangleq\tan^{-1}\left(  r_{\Im}/r_{\Re}\right)  $.
Then, $f\left(  r|\alpha,s_{m}\right)  $\ can be evaluated by substituting
(\ref{phase}) and (\ref{E-PDF-02}) into (\ref{E-PDF-01}).

Because evaluating the integral in (\ref{E-PDF-01}) is intractable, Von Mises
(Tikhonov or circular normal) distribution is used as approximation for
$f_{\theta}(\theta)$ in (\ref{phase}), which can be written as
\cite{Tikh-Rice}%
\begin{equation}
f(\theta)\approx\frac{1}{2\pi I_{0}\left(  2\sqrt{K\left(  K+1\right)
}\right)  }\exp\left(  2\sqrt{K\left(  K+1\right)  }\cos\left(  \theta
-\phi\right)  \right)  .
\end{equation}
Therefore, the integral in (\ref{E-PDF-01}) can be written as
\begin{equation}
f\left(  r|\alpha,s_{m}\right)  \approx G\left(  r\right)  \exp\left(
-\frac{\alpha^{2}s_{m}^{2}}{2\sigma_{n}^{2}}\right)  \mathcal{I}_{\theta
}\text{.}%
\end{equation}
where $G\left(  r\right)  $
\begin{equation}
G\left(  r\right)  =\frac{1}{2\pi I_{0}\left(  2\sqrt{K\left(  K+1\right)
}\right)  \sigma_{n}^{2}}\exp\left(  -\frac{\left\vert r\right\vert ^{2}%
}{2\sigma_{n}^{2}}\right)
\end{equation}
and
\begin{equation}
\mathcal{I}_{\theta}=\frac{1}{2\pi}\int_{-\pi}^{\pi}\exp\left(  2\sqrt
{K\left(  K+1\right)  }\cos\left(  \theta-\phi\right)  \right)  \exp\left(
\frac{\alpha s_{m}}{\sigma_{n}^{2}}\left\vert r\right\vert \cos\left(
\theta-\theta_{\text{r}}\right)  \right)  d\theta.\label{I-theta}%
\end{equation}
The factor $G\left(  r\right)  $ can be considered constant with respect to
the maximization process in (\ref{E-ACD-Optimum}), thus, it is more convenient
to separate it from the other terms. Moreover, for the special case where
$s_{m}=0$, the PDF$\ f\left(  r|\alpha\text{,}\theta\text{,}s_{m}=0\right)  $
is independent of $\alpha$\ and $\theta$. Thus,%
\begin{align}
f\left(  r|\alpha,\theta,s_{0}=0\right)   & =\frac{1}{2\pi\sigma_{n}^{2}}%
\exp\left[  -\frac{\left\vert r\right\vert ^{2}}{2\sigma_{n}^{2}}\right]
\nonumber\\
& =G\left(  r\right)  I_{0}\left(  2\sqrt{K\left(  K+1\right)  }\right)
\text{.}%
\end{align}
For $s_{m}\neq0$, evaluating the integral $\mathcal{I}_{\theta}$\ is actually
intractable due to the existence of $\phi$ and $\theta_{\mathrm{r}}$.
Moreover, $\theta_{\mathrm{r}}$ depends nonlinearly on $\theta$, which makes
it difficult to evaluate $\mathcal{I}_{\theta}$ even numerically. To overcome
this problem, we assume that $\theta_{\mathrm{r}}$ is independent of $\theta$,
consequently, the derived detector is near-optimal. Moreover, $\theta
_{\mathrm{r}}$ represents the phase of the received signal $r_{i}$, and hence
it can be computed directly at the receiver. The SER performance of the
near-optimal detector is expected to be close to the optimum at low SNRs
because the second exponent in (\ref{I-theta}) will be less significant, and
hence, the assumption that $\theta_{\mathrm{r}}$ and $\theta$ are independent
will not have substantial effect on the SER. On the contrary, at high SNRs,
the second exponent dominates the value of $\mathcal{I}_{\theta}$, and hence
the SER is expected to diverge from the optimum.

Based on the assumption that $\theta$ and $\theta_{\mathrm{r}}$ are
independent, the Gauss-Chebyshev quadrature integration rule can used as shown
in Appendix I to derive an approximate solution, which is given by
\begin{equation}
\mathcal{I}_{\theta}=\frac{1}{L}\sum\limits_{l=1}^{L}\exp\left[  \left(
\bar{K}\cos\left(  \phi\right)  +\tfrac{\cos\left(  \theta_{\text{r}}\right)
}{\sigma_{n}^{2}}\alpha s_{m}\left\vert r\right\vert \right)  \cos\left(
\varphi+\pi\tfrac{2l-1}{2L}\right)  +\left(  \bar{K}\sin\left(  \phi\right)
+\tfrac{\sin\left(  \theta_{\text{r}}\right)  }{\sigma_{n}^{2}}\alpha
s_{m}\left\vert r\right\vert \right)  \sin\left(  \varphi+\pi\tfrac{2l-1}%
{2L}\right)  \right] \label{Gauss-Cheb}%
\end{equation}
where $\bar{K}=2\sqrt{K\left(  K+1\right)  }$, $L$ is the quadrature order,
and
\[
\varphi=\tan^{-1}\left(  \frac{\bar{K}\sin\left(  \phi\right)  +\frac
{1}{\sigma_{n}^{2}}\sin\left(  \theta_{\text{r}}\right)  \alpha s_{m}%
\left\vert r\right\vert }{\bar{K}\cos\left(  \phi\right)  +\frac{1}{\sigma
_{n}^{2}}\cos\left(  \theta_{\text{r}}\right)  \alpha s_{m}\left\vert
r\right\vert }\right)  .
\]
Therefore, after applying the $\ln\left(  \cdot\right)  $ function and
dropping the common and constant terms, the optimum AC detector reduces to%
\begin{equation}
\hat{d}=\arg\min_{\tilde{s}_{m}\in\mathbb{S}}\text{ }\sum\limits_{i=1}%
^{N}\frac{\alpha_{i}^{2}\tilde{s}_{m}^{2}}{2\sigma_{n}^{2}}-\ln\tilde
{\mathcal{I}}_{\theta_{i}}\label{E-ACD-Opt-00}%
\end{equation}
where $\tilde{\mathcal{I}}_{\theta_{i}}=\mathcal{I}_{\theta_{i}}%
|_{s_{m}\rightarrow\tilde{s}_{m}}$. As can be noted from (\ref{E-ACD-Opt-00}),
the optimum AC detector has very high computational complexity induced by
$\mathcal{I}_{\theta_{i}}$, which makes it prohibitively expensive to
implement. Moreover, the detector requires the knowledge of the noise variance
$\sigma_{n}^{2}$ and the Rician fading parameter $K$. The received signal
phase can be computed directly from the received signal, $\theta_{\text{r}%
_{i}}\triangleq\tan^{-1}\left(  r_{i,\Im}/r_{i,\Re}\right)  .$

\subsubsection{Suboptimum AC Detector}

As can be noted from (\ref{I_theta}) in Appendix I, $B$ and $D$ dominates
$g\left(  \theta\right)  $ at high SNRs, i.e., $B\gg A\cos\left(  \phi\right)
$ and $D\gg C$. Thus, substituting $A=C=0$ in (\ref{I_theta}) yields%
\begin{align}
\mathcal{I}_{\theta}  & \approx\frac{1}{2\pi}\int_{-\pi}^{\pi}\exp\left(
B\alpha s_{m}\left\vert r\right\vert \cos\left(  \theta\right)  +D\alpha
s_{m}\left\vert r\right\vert \sin\left(  \theta\right)  \right)
d\theta\nonumber\\
& =\frac{1}{2\pi}\int_{-\pi}^{\pi}\exp\left(  \frac{\alpha s_{m}\left\vert
r\right\vert }{\sigma_{n}^{2}}\cos\left(  \theta_{\text{r}}\right)
\cos\left(  \theta\right)  +\frac{\alpha_{i}s_{m}\left\vert r\right\vert
}{\sigma_{n}^{2}}\sin\left(  \theta_{\text{r}}\right)  \sin\left(
\theta\right)  \right)  d\theta\nonumber\\
& =\frac{1}{2\pi}\int_{-\pi}^{\pi}\exp\left(  \frac{\alpha s_{m}\left\vert
r\right\vert }{\sigma_{n}^{2}}\left(  \cos\left(  \theta_{\text{r}}\right)
\cos\left(  \theta\right)  +\sin\left(  \theta_{\text{r}}\right)  \sin\left(
\theta\right)  \right)  \right)  d\theta\nonumber\\
& =\frac{1}{2\pi}\int_{-\pi}^{\pi}\exp\left(  \frac{\alpha s_{m}\left\vert
r\right\vert }{\sigma_{n}^{2}}\cos\left(  \theta-\theta_{\text{r}}\right)
\right)  d\theta\nonumber\\
& =\frac{1}{\pi}\int_{0}^{\pi}\exp\left(  \frac{\alpha s_{m}\left\vert
r\right\vert }{\sigma_{n}^{2}}\cos\left(  \theta\right)  \right)
d\theta\nonumber\\
& =I_{0}\left(  \frac{\alpha s_{m}\left\vert r\right\vert }{\sigma_{n}^{2}%
}\right)  .
\end{align}
It is worth noting that $\theta_{\text{r}}$\ does not affect the result of the
integral because it is only a phase shift. Therefore, the AC detector can be
expressed as%
\begin{equation}
\hat{d}=\arg\min_{\tilde{s}_{m}\in\mathbb{S}}\text{ }\sum\limits_{i=1}%
^{N}\frac{\alpha_{i}^{2}\tilde{s}_{m}^{2}}{2\sigma_{n}^{2}}-\ln\left[
I_{0}\left(  \frac{\alpha_{i}\tilde{s}_{m}\left\vert r_{i}\right\vert }%
{\sigma_{n}^{2}}\right)  \right]  .\label{E-ACD-subopt-00}%
\end{equation}
\bigskip The suboptimum AC detector described in (\ref{E-ACD-subopt-00}) is
similar to the optimum AC detector derived in \cite{Dweik}\ for Rayleigh
fading channels. Although this detector does not require knowledge of the
statistical channel information, it requires computing the Bessel function,
which incurs high complexity.

Another suboptimal detector can be obtained by directly substituting $A=0$ and
$C=0$ in (\ref{Gauss-Cheb}), therefore $\mathcal{I}_{\theta}$ can be
approximated as%
\begin{align}
\mathcal{I}_{\theta}  & \approx\frac{1}{L}\sum\limits_{l=1}^{L}\exp\left[
\frac{\alpha s_{m}\left\vert r\right\vert }{\sigma_{n}^{2}}\left(  \cos\left(
\theta_{\text{r}}\right)  \cos\left(  \varphi+\pi\tfrac{2l-1}{2L}\right)
+\sin\left(  \theta_{\text{r}}\right)  \sin\left(  \varphi+\pi\tfrac{2l-1}%
{2L}\right)  \right)  \right] \nonumber\\
& =\frac{1}{L}\sum\limits_{l=1}^{L}\exp\left[  \frac{\alpha s_{m}\left\vert
r\right\vert }{\sigma_{n}^{2}}\left(  \cos\left(  -\pi\tfrac{2l-1}{2L}\right)
\right)  \right] \nonumber\\
& =\frac{1}{L}\sum\limits_{l=1}^{L}\exp\left[  \frac{\alpha s_{m}\left\vert
r\right\vert }{\sigma_{n}^{2}}\cos\left(  \pi\tfrac{2l-1}{2L}\right)  \right]
.
\end{align}
Interestingly, this approach does not contain the Bessel function.

\subsubsection{Heuristic ACD (HACD)}

Although the two suboptimum AC detectors derived above\ are less complex than
the optimum AC detector, evaluating the Bessel and exponential functions is
necessary to calculate the decision metric. Therefore, the heuristic detector
presented in \cite{Dweik} is considered to reduce the complexity even further.
The heuristic detector is given by%
\begin{equation}
\hat{d}=\arg\min_{\tilde{s}_{m}\in\mathbb{S}}\text{ }\left[  \zeta-\tilde
{s}_{m}^{2}\right]  ^{2},\label{HeD}%
\end{equation}
where $\zeta$ is the combined signal from the $N$ antennas, which is given by%
\begin{equation}
\zeta=\frac{\left\vert r\right\vert _{\Sigma}^{2}}{\sum_{i=1}^{N}\alpha
_{i}^{2}},
\end{equation}
where $\left\vert r\right\vert _{\Sigma}^{2}=\sum_{i=1}^{N}\left\vert
r_{i}\right\vert ^{2}$. For SER analysis, it is more convenient to express
(\ref{HeD}) as
\begin{equation}
\hat{d}=\left\{
\begin{array}
[c]{cc}%
s_{0}\text{,} & \ \ \ 0<\zeta<\eta_{0,1}\\
s_{1}\text{,} & \eta_{0,1}<\zeta<\eta_{1,2}\\
\vdots & \vdots\\
s_{M-1}\text{,} & \eta_{M-1,M-2}<\zeta<\infty
\end{array}
\right.  ,\label{E-d-hat}%
\end{equation}
where $\eta_{i,j}$'s are the detection thresholds and given by%
\begin{equation}
\eta_{i,j}=\frac{s_{i}^{2}+s_{j}^{2}}{2}.\label{E-Thresh_Heu}%
\end{equation}
In the following two sections, two different approaches are presented to
evaluate the SER analytically.

\section{Approach I: SER Analysis of the Heuristic AC Detector}

Based on (\ref{E-d-hat}), the SER $P_{e}$ for the heuristic AC detector can be
written as,%
\begin{align}
P_{e}  & =1-\frac{1}{M}\left(  \int_{0}^{\eta_{0,1}}f\left(  \zeta
|E_{0}\right)  d\zeta+\int_{\eta_{M-2,M-1}}^{\infty}f\left(  \zeta
|E_{M-1}\right)  d\zeta+\sum_{m=1}^{M-2}\int_{\eta_{m,m-1}}^{\eta_{m,m+1}%
}f\left(  \zeta|E_{m}\right)  d\zeta\right) \nonumber\\
& =1-\frac{1}{M}\left(  F_{\zeta}\left(  \eta_{0,1}|E_{0}\right)  +1-F_{\zeta
}\left(  \eta_{M-2,M-1}|E_{M-1}\right)  +\sum_{m=1}^{M-2}F_{\zeta}\left(
\eta_{m,m+1}|E_{m}\right)  -F_{\zeta}\left(  \eta_{m,m-1}|E_{m}\right)
\right)
\end{align}
where $F_{\zeta}$ is the cumulative distribution function (CDF), which
typically can be evaluated as
\begin{equation}
F_{\zeta}\left(  \zeta|E_{m}\right)  =\int_{0}^{\zeta}\int_{0}^{\infty}%
\cdots\int_{0}^{\infty}f\left(  \zeta|\mathbf{\alpha},E_{m}\right)  f\left(
\mathbf{\alpha}\right)  d\mathbf{\alpha}\text{ }d\zeta\text{.}\label{E-CDF-00}%
\end{equation}
However, \textup{the} $N$-fold integral in (\ref{E-CDF-00}) can be
substantially simplified by noting that $f\left(  \zeta|\mathbf{\alpha}%
,E_{m}\right)  $ is actually a function of $\sum_{i=1}^{N}\alpha_{i}%
^{2}\triangleq x.$ Consequently, the integral reduces to
\begin{equation}
F_{\zeta}\left(  \zeta|E_{m}\right)  =\int_{0}^{\zeta}\int_{0}^{\infty
}f\left(  \zeta|x,E_{m}\right)  f\left(  x\right)  dx\text{ }d\zeta
\text{,}\label{E-CDF-01}%
\end{equation}

The PDF of $x$ can be derived by noting that $h_{i}\sim\mathcal{CN}\left(
m_{h},2\sigma_{h}^{2}\right)  $, thus $f\left(  x\right)  $ is noncentral
Chi-squared%
\begin{equation}
f\left(  x\right)  =\frac{\exp\left(  -\frac{\lambda}{2}\right)  }{2\sigma
_{h}^{2}}\exp\left(  -\frac{x}{2\sigma_{h}^{2}}\right)  \left(  \frac
{x}{\lambda\sigma_{h}^{2}}\right)  ^{0.5\left(  N-1\right)  }I_{N-1}\left(
\sqrt{\frac{\lambda}{\sigma_{h}^{2}}x}\right)  ,\label{PDF-x}%
\end{equation}
where%
\begin{align}
\lambda & =\sum_{i=1}^{N}\mu_{h_{I}}^{2}+\mu_{h_{Q}}^{2}\nonumber\\
& =\sum_{i=1}^{N}\frac{\mu_{h}^{2}\cos^{2}\phi}{\sigma_{h}^{2}}+\frac{\mu
_{h}^{2}\sin^{2}\phi}{\sigma_{h}^{2}}\nonumber\\
& =2KN.
\end{align}
and the complete derivation can be found in Appendix II. To derive $f\left(
\zeta|x,E_{m}\right)  $ of the Heuristic detector (\ref{HeD}), the decision
variable can be written as%
\begin{align}
\zeta & =\frac{\left\vert r\right\vert _{\Sigma}^{2}}{\sum_{i=1}^{N}\alpha
_{i}^{2}}\nonumber\\
& =\frac{1}{\sum_{i=1}^{N}\alpha_{i}^{2}}\sum_{i=1}^{N}\left\vert
r_{i}\right\vert ^{2}\nonumber\\
& =\frac{1}{\sum_{i=1}^{N}\alpha_{i}^{2}}\underset{r_{T}}{\underbrace{\sum
_{i=1}^{N}\left\vert r_{i,\Re}+jr_{i,\Im}\right\vert ^{2}}},\label{He-D}%
\end{align}
where $r_{i,\Re}=\alpha_{i}s_{m}\cos\left(  \theta_{i}\right)  +n_{i,\Re}$ and
$r_{i,\Im}=\alpha_{i}s_{m}\sin\left(  \theta_{i}\right)  +n_{i,\Im}$. By
noting that $r_{i,\Re}$ and $r_{i,\Im}$ are mutually independent $\forall i$,
and the PDF for each of which is conditionally Gaussian, i.e., $f\left(
r_{i,\Re}|\alpha_{i},s_{m},\cos\left(  \theta_{i}\right)  \right)
\sim\mathcal{N}\left(  \alpha_{i}s_{m}\cos\left(  \theta_{i}\right)
,\sigma_{n}^{2}\right)  $ and $f\left(  r_{i,\Im}|\alpha_{i},s_{m},\sin\left(
\theta_{i}\right)  \right)  \sim\mathcal{N}\left(  \alpha_{i}s_{m}\sin\left(
\theta_{i}\right)  ,\sigma_{n}^{2}\right)  $. Therefore, $f\left(  \left\vert
r_{i}\right\vert ^{2}|\alpha_{i},s_{m},\theta_{i}\right)  $is conditionally
noncentral Chi-squared with two degrees of freedom and noncentrality factor
$\lambda_{i}$, i.e., $f\left(  \left\vert r_{i}\right\vert ^{2}|\alpha
_{i},s_{m},\theta_{i}\right)  \sim\chi^{2}\left(  2,\lambda_{i}\right)  $,
where
\begin{align}
\lambda_{i}  & =\left[  \alpha_{i}s_{m}\cos\left(  \theta_{i}\right)  \right]
^{2}+\left[  \alpha_{i}s_{m}\cos\left(  \theta_{i}\right)  \right]
^{2}\nonumber\\
& =\alpha_{i}^{2}s_{m}^{2}.
\end{align}
Therefore, $f\left(  \left\vert r_{i}\right\vert ^{2}|\alpha_{i},s_{m}%
,\theta_{i}\right)  =f\left(  \left\vert r_{i}\right\vert ^{2}|\alpha
_{i},s_{m}\right)  \sim\chi^{2}\left(  2,\lambda_{i}\right)  $. Therefore,
$f\left(  \zeta|x\right)  $ can be written as%
\begin{equation}
f\left(  \zeta|x\mathbf{,}E_{m}\right)  =\frac{x}{2\sigma_{n}^{2}}\left(
\frac{\zeta}{E_{m}}\right)  ^{0.5\left(  N-1\right)  }\exp\left(
-x\frac{\zeta+E_{m}}{2\sigma_{n}^{2}}\right)  I_{N-1}\left(  \frac{x}%
{\sigma_{n}^{2}}\sqrt{E_{m}\zeta}\right)  \text{, }m>0.\label{f-zeta-em}%
\end{equation}

For the case of $m=0$, $r_{i,\Re}=n_{i,\Re}\sim\mathcal{N}\left(  0,\sigma
_{n}^{2}\right)  $ and $r_{i,\Im}=n_{i,\Im}\sim\mathcal{N}\left(  0,\sigma
_{n}^{2}\right)  $, thus, $f\left(  \left\vert r_{i}\right\vert ^{2}%
|\alpha_{i},d_{m},\theta_{i}\right)  =f\left(  \left\vert r_{i}\right\vert
^{2}\right)  $\ which has exponential PDF with parameter $\beta=2\sigma
_{n}^{2}$. Consequently, the PDF of the sum of $N$ iid exponential random
variables is Erlang distribution, i.e.,%
\begin{equation}
f\left(  \zeta|x\right)  =\frac{\lambda_{0}^{N}}{\left(  N-1\right)  !}%
x^{N}\text{ }\zeta^{N-1}\exp\left(  -\lambda_{0}\zeta x\right)  \text{, }m=0.
\end{equation}
where $\lambda_{0}=\frac{1}{2\sigma_{n}^{2}}.$

\subsection{Evaluating $F_{\zeta}\left(  \zeta|E_{m}\right)  $, $m>0$}

To simplify the analysis, we replace the Bessel function in (\ref{PDF-x}) by
its series expansion \cite{Table}, which gives%
\begin{equation}
I_{N-1}\left(  \sqrt{\frac{\lambda}{\sigma_{h}^{2}}x}\right)  =\sum
_{l=0}^{\infty}\frac{1}{l!\Gamma\left(  l+N\right)  }\left(  \frac{\lambda
}{4\sigma_{h}^{2}}\right)  ^{\frac{2l+N-1}{2}}\left(  x\right)  ^{\frac
{2l+N-1}{2}}.\label{apprx1}%
\end{equation}
Although the same series expansion can be used to represent the Bessel
function in (\ref{f-zeta-em}), the argument of the Bessel function
$\sqrt{\frac{E_{m}\text{ }\zeta}{\sigma_{n}^{2}}}x\gg1$ for most typical
values of $\sigma_{n}^{2}$. Therefore, the following approximation can be used
\cite{Stegun},%
\begin{equation}
I_{N-1}\left(  \frac{x}{\sigma_{n}^{2}}\sqrt{E_{m}\zeta}\right)  \simeq
\frac{\exp\left(  \tfrac{x\sqrt{E_{m}\text{ }\zeta}}{\sigma_{n}^{2}}\right)
}{\sqrt[4]{E_{m}\text{ }\zeta}\sqrt{\frac{2\pi}{\sigma_{n}^{2}}x}}\left(
1+\sum_{q=1}^{Q}\left(  \frac{\left(  -1\right)  ^{q}}{x^{q}}\frac{\prod
_{k=1}^{q}\left[  4\left(  N-1\right)  ^{2}-\left(  2k-1\right)  ^{2}\right]
}{q!8^{q}\left(  \frac{\sqrt{E_{m}\zeta}}{\sigma_{n}^{2}}\right)  ^{q}%
}\right)  \right)  .\label{apprx2}%
\end{equation}
Using (\ref{apprx1}) and (\ref{apprx2}), and noting that
\begin{equation}
f\left(  \zeta|E_{m}\right)  =\int_{0}^{\infty}f\left(  \zeta|x,E_{m}\right)
\text{ }f\left(  x\right)  dx,
\end{equation}
the PDF $f\left(  \zeta|E_{m}\right)  $ after evaluating the integral, as
depicted in Appendix II, is given by%
\begin{equation}
f\left(  \zeta|E_{m}\right)  =C\text{ }\left(  \sum_{l=0}^{\infty}A_{l}%
\zeta\text{ }^{0.5N-0.75}\text{ }C_{\zeta}^{-\left(  l+N+0.5\right)  }%
+\sum_{q=1}^{Q}\sum_{l=0}^{L}B_{q}^{l}\text{ }\zeta^{0.5N-0.5q-0.75}\text{
}C_{\zeta}^{-\left(  N-q+l+0.5\right)  }\right)  \text{, }m>0
\end{equation}
\bigskip where the variables $C$, $C_{\zeta}$, $A_{l}$ and $B_{q}^{l}$ are
given by%
\begin{equation}
C_{\zeta}\triangleq\frac{\zeta+E_{m}}{2\sigma_{n}^{2}}+\frac{\sigma_{h}^{2}%
}{2}-\frac{\sqrt{E_{m}\text{ }\zeta}}{\sigma_{n}^{2}}%
\end{equation}%
\begin{equation}
C=\frac{\exp\left(  -\frac{\lambda}{2}\right)  }{4\sigma_{n}^{2}\sigma_{h}%
^{2}}\left(  \frac{1}{E_{m}\sigma_{h}^{2}\lambda}\right)  ^{0.5\left(
N-1\right)  }\frac{1}{\sqrt{2\pi\frac{\sqrt{E_{m}}}{\sigma_{n}^{2}}}}%
\end{equation}%
\begin{equation}
A_{l}=\frac{\Gamma\left(  l+N+0.5\right)  }{l!\Gamma\left(  l+N\right)
}\left(  \frac{\lambda}{4\sigma_{h}^{2}}\right)  ^{\frac{2l+N-1}{2}}%
\end{equation}%
\begin{equation}
B_{q}^{l}=\frac{\Gamma\left(  N-q+l+0.5\right)  }{l!\Gamma\left(  l+N\right)
}\left(  \frac{\lambda}{4\sigma_{h}^{2}}\right)  ^{\frac{2l+N-1}{2}}\left(
\left(  -1\right)  ^{q}\frac{\prod_{k=1}^{q}\left[  4\left(  N-1\right)
^{2}-\left(  2k-1\right)  ^{2}\right]  }{q!8^{q}\left(  \frac{\sqrt{E_{m}}%
}{\sigma_{n}^{2}}\right)  ^{q}}\right)  ,
\end{equation}
where $\Gamma\left(  \cdot\right)  $ is the Gamma function \cite{Table}.

Finally, the CDF $F_{\zeta}\left(  \zeta|E_{m}\right)  $, $m>0$ can be
evaluated as%
\begin{align}
F_{\zeta}\left(  \zeta|E_{m}\right)   & =\int_{0}^{\zeta}f\left(  \acute
{\zeta}|E_{m}\right)  d\acute{\zeta}\nonumber\\
& =C\left(  \sum_{l=0}^{\infty}A_{l}\mathcal{I}_{A}^{l}+\sum_{q=1}^{Q}%
\sum_{l=0}^{\infty}B_{m}^{l}\mathcal{I}_{B}^{q,l}\right)  \text{, }m>0,
\end{align}
where $\mathcal{I}_{A}^{l}$\ and $\mathcal{I}_{B}^{q,l}$\ are given by%
\begin{align}
\mathcal{I}_{A}^{l}  & =2\int_{0}^{\sqrt{\zeta}}y^{2\left(  0.5N-0.75\right)
+1}\left(  \frac{y^{2}}{2\sigma_{n}^{2}}-\frac{\sqrt{E_{m}}}{\sigma_{n}^{2}%
}y+c\right)  ^{-\left(  l+N+0.5\right)  }dy\label{intga}\\
\mathcal{I}_{B}^{q,l}  & =2\int_{0}^{\sqrt{\zeta}}y^{2\left(
0.5N-0.5q-0.75\right)  +1}\left(  \frac{y^{2}}{2\sigma_{n}^{2}}-\frac
{\sqrt{E_{m}}}{\sigma_{n}^{2}}y+c\right)  ^{-\left(  N-q+l+0.5\right)
}dy,\label{intgb}%
\end{align}
where $c=\frac{1}{2\sigma_{h}^{2}}+\frac{E_{m}}{2\sigma_{n}^{2}}$.

\subsection{Evaluating $F_{\zeta}\left(  \zeta|E_{m}\right)  $, $m=0$}

Averaging $\textup{f}\left(  \zeta|x,E_{m}\right)  $ over $\textup{f}\left(
x\right)  $\ using series expansion of the Bessel function using
(\ref{apprx1})\ yields%
\begin{align}
f\left(  \zeta|E_{m}\right)   & =\int_{0}^{\infty}\textup{f}\left(
\zeta|x,E_{m}\right)  \text{ }f\left(  x\right)  dx\nonumber\\
& =C_{0}\text{ }\zeta^{N-1}\sum_{l=0}^{\infty}\frac{\Gamma\left(  2N+l\right)
}{l!\Gamma\left(  l+N\right)  }\left(  \frac{\lambda_{0}}{4\sigma_{h}^{2}%
}\right)  ^{\frac{2l+N-1}{2}}\left(  \lambda_{0}\zeta+\frac{1}{2\sigma_{h}%
^{2}}\right)  ^{-\left(  2N+l\right)  }\text{, }m=0
\end{align}
where $C_{0}$ is given by%
\begin{equation}
C_{0}=\frac{\lambda_{0}^{N}}{\left(  N-1\right)  !}\frac{1}{2\sigma_{h}^{2}%
}\left(  \frac{1}{\lambda_{0}\sigma_{h}^{2}}\right)  ^{0.5\left(  N-1\right)
}\exp\left(  -\frac{\lambda_{0}}{2}\right)  .
\end{equation}
The complete derivation of $f\left(  \zeta|E_{m}\right)  $\ is given in
Appendix III.

Finally, the CDF of $\zeta$ can be evaluated as%
\begin{align}
F_{\zeta}\left(  \zeta|E_{m}\right)   & =\int_{0}^{\zeta}f\left(  \acute
{\zeta}|E_{m}\right)  d\acute{\zeta}\nonumber\\
& =C_{0}\sum_{l=0}^{\infty}\frac{\Gamma\left(  2N+l\right)  }{l!\Gamma\left(
l+N\right)  }\left(  \frac{\lambda_{0}}{4\sigma_{h}^{2}}\right)
^{\frac{2l+N-1}{2}}\int_{0}^{\zeta}\acute{\zeta}^{N-1}\left(  \lambda
_{0}\text{ }\acute{\zeta}+\frac{1}{2\sigma_{h}^{2}}\right)  ^{-\left(
2N+l\right)  }d\acute{\zeta}\text{, }m=0.\label{intg0}%
\end{align}
It should be noticed that integrals of the form given in (\ref{intga}),
(\ref{intgb}) and (\ref{intg0}) can be solved in recursive manner according to
\cite[2.17, page 78]{Table}.

\section{Approach II: Symbol Error Rate Analysis}

In the first approach, we used the conditional PDF $f\left(  \zeta
|x,E_{m}\right)  $ to derive the unconditional PDF $f\left(  \zeta
|E_{m}\right)  $, which was used to derive the unconditional CDF. In this
part, we derive the conditional CDF $F_{\zeta}\left(  \zeta|x,E_{m}\right)  $
from the conditional PDF $f\left(  \zeta|x,E_{m}\right)  $, and then we derive
the unconditional CDF $F_{\zeta}\left(  \zeta|E_{m}\right)  .$

By noting that $\zeta=\frac{\left\vert r\right\vert _{\Sigma}^{2}}{x}$, then
the conditional PDF $f\left(  \zeta|x,E_{m}\right)  $ follows an noncentral
Chi-squared with $2N$ degrees of freedom, noncentrality parameter $E_{m}$, and
the variance of Gaussian components is $\frac{\sigma_{n}^{2}}{x}$, i.e.,
$f\left(  \zeta|x,E_{m}\right)  \sim\chi^{2}\left(  2N,s_{m}^{2}\right)
=\chi^{2}\left(  2N,E_{m}\right)  $. Thus, $F_{\zeta}\left(  \zeta
|x,E_{m}\right)  $ \cite{porakis} can be defined as%
\begin{equation}
F_{\zeta}\left(  \zeta|x,E_{m}\right)  =1-{Q}_{N}\left(  \sqrt{\frac{xE_{m}%
}{\sigma_{n}^{2}}},\sqrt{\frac{x\zeta}{\sigma_{n}^{2}}}\right)
.\label{E-CDF-MQ-00}%
\end{equation}
The CDF of $\zeta$ given $E_{m}$ can be calculated by averaging the
conditional CDF $F\left(  \zeta|x,E_{m}\right)  $ over the distribution of $x$
which is given by
\begin{equation}
F\left(  \zeta|E_{m}\right)  =\int_{0}^{\infty}F\left(  \zeta|x,E_{m}\right)
f(x)dx,
\end{equation}
where $f(x)$ is given in (\ref{PDF-x}).

\subsection{The CDF $F_{\zeta}(\zeta|E_{m})$ for $\zeta<E_{m}$}

The series representation of the generalized Marcum Q-function is given in
\cite{andras} as
\begin{equation}
Q_{v}(a,b)=1-\sum_{n=0}^{\infty}(-1)^{n}\exp\left(  -\frac{a^{2}}{2}\right)
\frac{L_{n}^{(v-1)}\left(  \frac{a^{2}}{2}\right)  }{\Gamma(v+n+1)}\left(
\frac{b^{2}}{2}\right)  ^{n+v},\;\;\left\{  a,v\right\}  >0\text{ and }b\geq0
\end{equation}
where $L_{n}^{(\alpha)}(x)=\sum_{k=0}^{n}\frac{\Gamma(n+\alpha+1)}%
{\Gamma(k+\alpha+1)\Gamma(n-k+1)}\frac{(-x)^{k}}{k!}$ is the generalized
Laguerre polynomial of degree $n$ and order $\alpha$. Using this formula, the
conditional CDF $F\left(  \zeta|x,E_{m}\right)  $ can be rewritten as%
\begin{equation}
F_{\zeta}\left(  \zeta|x,E_{m}\right)  =\sum_{n=0}^{\infty}\sum_{k=0}%
^{n}{\binom{N+n-1}{N+k-1}}\frac{(-1)^{n+k}}{(N+n)!k!}\left(  \frac{\zeta
}{2\sigma_{n}^{2}}\right)  ^{N+n}\left(  \frac{E_{m}}{2\sigma_{n}^{2}}\right)
^{k}x^{N+n+k}\exp\left(  -\frac{E_{m}}{2\sigma_{n}^{2}}x\right)  .
\end{equation}
Then, the CDF $F_{\zeta}(\zeta|E_{m})$ can be computed as%
\begin{equation}
F_{\zeta}(\zeta|E_{m})=\sum_{n=0}^{\infty}\sum_{k=0}^{n}{\binom{N+n-1}{N+k-1}%
}\frac{(-1)^{n+k}}{(N+n)!k!}\left(  \frac{\zeta}{2\sigma_{n}^{2}}\right)
^{N+n}\left(  \frac{E_{m}}{2\sigma_{n}^{2}}\right)  ^{k}%
\underset{A}{\underbrace{\int_{0}^{\infty}x^{N+n+k}e^{-\frac{E_{m}}%
{2\sigma_{n}^{2}}x}f(x)dx}}.\label{cdf_wA}%
\end{equation}
The integration $A$ can be solved by\ \cite[Eq. 2.15.5]{prudnikov}%
\begin{align}
A  & =2\bar{K}\text{e}^{-NK}\left(  \frac{\bar{K}}{NK}\right)  ^{\frac{N-1}%
{2}}\int_{0}^{\infty}\alpha^{2(N+n+k)+N}\exp\left(  -\left(  \frac{E_{m}%
}{2\sigma_{n}^{2}}+\bar{K}\right)  \alpha^{2}\right)  I_{N-1}\left(
2\sqrt{NK\bar{K}}\alpha\right)  d\alpha\nonumber\\
& =\text{e}^{-NK}\bar{K}^{N}\frac{\Gamma(n+k+2N)}{\Gamma(N)}\left(
\frac{E_{m}}{2\sigma_{n}^{2}}+\bar{K}\right)  ^{-(n+k+2N)}\text{ }{_{1}F_{1}%
}\left(  n+k+2N;N;\frac{NK(1+K)}{1+K+\frac{\Omega}{2\sigma_{n}^{2}}E_{m}%
}\right)  ,
\end{align}
which can be simplified to%
\begin{align}
A  & =\frac{e^{-\frac{\lambda}{2}}}{\sigma_{h}^{2}}\left(  \frac{1}%
{\lambda\sigma_{h}^{2}}\right)  ^{\frac{N-1}{2}}\int_{0}^{\infty}%
\alpha^{2(N+n+k)+N}\exp\left(  -\left(  \frac{E_{m}}{2\sigma_{n}^{2}}+\frac
{1}{2\sigma_{h}^{2}}\right)  \alpha^{2}\right)  I_{N-1}\left(  \sqrt
{\frac{\lambda}{\sigma_{h}^{2}}}\alpha\right)  d\alpha\nonumber\\
& =e^{-\frac{\lambda}{2}}\left(  \frac{1}{2\sigma_{h}^{2}}\right)  ^{N}%
\frac{\Gamma(n+k+2N)}{\Gamma(N)}\left(  \frac{E_{m}}{2\sigma_{n}^{2}}+\frac
{1}{2\sigma_{h}^{2}}\right)  ^{-(n+k+2N)}\text{ }{_{1}F_{1}}\left(
n+k+2N;N;\frac{\lambda}{2\left(  1+\frac{\sigma_{h}^{2}}{\sigma_{n}^{2}}%
E_{m}\right)  }\right)  ,
\end{align}
where $\bar{K}=\frac{1+K}{\Omega}$ and ${_{1}F_{1}}(a;b;z)$ is the confluent
hypergeometric function of the first kind.

Consequently, by substituting $A$ into (\ref{cdf_wA}) and applying some
manipulations, the CDF can be expressed as
\begin{multline}
F_{\zeta}(\zeta|E_{m})=\sum_{n=0}^{\infty}\sum_{k=0}^{n}{\binom{N+n-1}{N+k-1}%
}\frac{(-1)^{n+k}(n+k+2N-1)!}{(N-1)!(N+n)!k!}\left(  \frac{\zeta}{E_{m}%
}\right)  ^{n+N}\left(  K_{m}\text{e}^{-K}\right)  ^{N}\\
\times\frac{\left(  \frac{\Omega}{2\sigma_{n}^{2}}\right)  ^{n+k+N}}{\left(
\frac{\Omega}{2\sigma_{n}^{2}}+K_{m}\right)  ^{n+k+2N}}{_{1}F_{1}}\left(
n+k+2N;N;\frac{NKK_{m}}{\frac{\Omega}{2\sigma_{n}^{2}}+K_{m}}\right)
,\label{cdf_zeta}%
\end{multline}
where $K_{m}=\frac{1+K}{E_{m}}$. Note that $\left(  \frac{\zeta}{E_{m}%
}\right)  ^{n}$ in the CDF expression is converged only with $\frac{\zeta
}{E_{m}}<1$ when $n$ goes to infinity. Therefore, it will be used to calculate
the error probability for $\Pr\left(  \zeta\leq\frac{E_{m}+E_{m-1}}{2}\right)
$.

For high SNR regime, when $\frac{\Omega}{2\sigma_{n}^{2}}\rightarrow\infty$,
(\ref{cdf_zeta}) can be expressed as%
\[
\lim_{\frac{\Omega}{2\sigma_{n}^{2}}\rightarrow\infty}\left(  \frac
{\frac{\Omega}{2\sigma_{n}^{2}}}{\frac{\Omega}{2\sigma_{n}^{2}}+K_{m}}\right)
^{n+k+N}{_{1}F_{1}}\left(  n+k+2N;N;\frac{NKK_{m}}{\frac{\Omega}{2\sigma
_{n}^{2}}+K_{m}}\right)  =1.
\]
Then, (\ref{cdf_zeta}) can be simplified as
\begin{gather}
F^{\infty}(\zeta|E_{m})=\left(  \frac{K_{m}e^{-K}}{\frac{\Omega}{2\sigma
_{n}^{2}}+K_{m}}\right)  ^{N}\sum_{n=0}^{\infty}\left(  \frac{\zeta}{E_{m}%
}\right)  ^{n+N}\sum_{k=0}^{n}{\binom{N+n-1}{N+k-1}}\frac{(-1)^{n+k}%
(n+k+2N-1)!}{(N-1)!(N+n)!k!}\nonumber\\
\overset{\text{(a)}}{=}\left(  \frac{K_{m}e^{-K}}{\frac{\Omega}{2\sigma
_{n}^{2}}+K_{m}}\right)  ^{N}\sum_{n=0}^{\infty}\left(  \frac{\zeta}{E_{m}%
}\right)  ^{n+N}\frac{(2N+n-1)!}{n!N!(N-1)!}\nonumber\\
\overset{\text{(b)}}{=}{\binom{2N-1}{N}}\left(  \frac{K_{m}e^{-K}}%
{\frac{\Omega}{2\sigma_{n}^{2}}+K_{m}}\frac{\frac{\zeta}{E_{m}}}{\left(
\frac{\zeta}{E_{m}}-1\right)  ^{2}}\right)  ^{N},\;(\zeta<E_{m})\label{asymA}%
\end{gather}
where (a) comes from
\[
\sum_{k=0}^{n}(-1)^{n+k}{\binom{N+n-1}{N+k-1}}\frac{(n+k+2N-1)!}{k!}%
=\frac{(n+2N-1)!(n+N)!}{n!N!}%
\]
and (b) comes from
\[
\sum_{n=0}^{\infty}a^{n}\frac{(n+2N-1)!}{n!}=(a-1)^{-2N}(2N-1)!\text{, }a<1.
\]

\subsection{The CDF $F_{\zeta}(\zeta|E_{m})$ for $\zeta>E_{m}$}

To evaluate the required CDF, the following relation between $Q_{m}(a,b)$ and
$Q_{m}(b,a)$ is applied.%
\begin{equation}
Q_{m}(a,b)+Q_{m}(b,a)=1+e^{-\frac{a^{2}+b^{2}}{2}}\sum_{k=1-m}^{m-1}\left(
\frac{a}{b}\right)  ^{k}I_{k}(ab).
\end{equation}
Using this transformation, the CDF $F_{\zeta}(\zeta|x,E_{m})$ for $\zeta
>E_{m}$ can be now rewritten as%
\begin{equation}
F_{\zeta}(\zeta|x,E_{m})={Q}_{N}\left(  \sqrt{\frac{x\zeta}{\sigma_{n}^{2}}%
},\sqrt{\frac{xE_{m}}{\sigma_{n}^{2}}}\right)  -\exp\left(  -\frac{\zeta
+E_{m}}{2\sigma_{n}^{2}}x\right)  \sum_{k=1-N}^{N-1}\left(  \sqrt{\frac{E_{m}%
}{\zeta}}\right)  ^{k}I_{k}\left(  \frac{\sqrt{E_{m}\zeta}}{\sigma_{n}^{2}%
}x\right)  .
\end{equation}
The CDF of $F_{\zeta}(\zeta|E_{m})$ can be calculated as%
\begin{equation}
F_{\zeta}(\zeta|E_{m})=\underset{B}{\underbrace{\int_{0}^{\infty}{Q}%
_{m}\left(  \sqrt{\tfrac{x\zeta}{\sigma_{n}^{2}}},\sqrt{\tfrac{xE_{m}}%
{\sigma_{n}^{2}}}\right)  f(x)dx}}-\underset{\acute{C}}{\underbrace{\sum
_{k=1-N}^{N-1}\left(  \sqrt{\tfrac{E_{m}}{\zeta}}\right)  ^{k}\int_{0}%
^{\infty}\exp\left(  -\tfrac{\zeta+E_{m}}{2\sigma_{n}^{2}}x\right)
I_{k}\left(  \tfrac{\sqrt{E_{m}\zeta}}{\sigma_{n}^{2}}x\right)  f(x)dx}%
},\label{E-CDF-Kihong-00}%
\end{equation}
where integrals $B$ and $\acute{C}$ can be evaluated as shown below in
(\ref{cdf_B}) and (\ref{cdf_C}), respectively. Integration in $B$ can be
calculated similarly as (\ref{cdf_zeta}) which is given by%
\begin{multline}
B=1-\sum_{n=0}^{\infty}\sum_{k=0}^{n}{\binom{N+n-1}{N+k-1}}\frac
{(-1)^{n+k}(n+k+2N-1)!}{(N-1)!(N+n)!k!}\left(  \frac{E_{m}}{\zeta}\right)
^{n+N}\left(  K_{\zeta}\text{\textrm{e}}^{-K}\right)  ^{N}\frac{\left(
\frac{\Omega}{2\sigma_{n}^{2}}\right)  ^{n+k+N}}{\left(  \frac{\Omega}%
{2\sigma_{n}^{2}}+K_{\zeta}\right)  ^{n+k+2N}}\\
\times{_{1}F_{1}}\left(  n+k+2N;N;\frac{NKK_{\zeta}}{\frac{\Omega}{2\sigma
_{n}^{2}}+K_{\zeta}}\right)  ,\label{cdf_B}%
\end{multline}
where $K_{\zeta}=\frac{1+K}{\zeta}$.

Using the series representation of modified Bessel function of the first kind,
i.e.,
\[
I_{v}(z)=\left(  \frac{z}{2}\right)  ^{v}\sum_{n=0}^{\infty}\frac{\left(
z^{2}/4\right)  ^{n}}{n!(n+v)!}%
\]
and noting that $I_{v}(z)=I_{-v}(z)$, the integral $\acute{C}$ can be
calculated as%
\begin{align}
\acute{C}  & =\sum_{n=0}^{\infty}\sum_{k=1-N}^{N-1}\left(  \frac{E_{m}}{\zeta
}\right)  ^{k}\frac{\left(  \frac{\sqrt{E_{m}\zeta}}{2\sigma_{n}^{2}}\right)
^{2n+|k|}}{n!(n+|k|)!}\int_{0}^{\infty}x^{2n+|k|}\exp\left(  -\frac
{\zeta+E_{m}}{2\sigma_{n}^{2}}x\right)  f(x)dx\nonumber\\
& =\sum_{n=0}^{\infty}\sum_{k=1-N}^{N-1}\left(  \frac{E_{m}}{\zeta}\right)
^{k}\frac{\left(  \frac{\sqrt{E_{m}\zeta}}{2\sigma_{n}^{2}}\right)  ^{2n+|k|}%
}{n!(n+|k|)!}\frac{(2n+|k|+N-1)!}{(N-1)!}e^{-NK}\left(  \frac{1}{2\sigma
_{h}^{2}}\right)  ^{N}\nonumber\\
& \times\left(  \frac{\zeta+E_{m}}{2\sigma_{n}^{2}}+\frac{1}{2\sigma_{h}^{2}%
}\right)  ^{-(2n+|k|+N)}{_{1}F_{1}}\left(  2n+|k|+N;N;\frac{\lambda}{2\left(
1+\frac{\sigma_{h}^{2}}{\sigma_{n}^{2}}(E_{m}+\zeta)\right)  }\right)
\nonumber\\
& =\sum_{n=0}^{\infty}\sum_{k=1-N}^{N-1}\left(  \frac{\sqrt{E_{m}\zeta}}%
{E_{m}+\zeta}\right)  ^{2n+|k|}\left(  \frac{E_{m}}{\zeta}\right)  ^{k}%
\frac{(2n+|k|+N-1)!}{n!(n+|k|)!(N-1)!}\left(  e^{-K}K_{m\zeta}\right)
^{N}\frac{\left(  \frac{\Omega}{2\sigma_{n}^{2}}\right)  ^{2n+|k|}}{\left(
\frac{\Omega}{2\sigma_{n}^{2}}+K_{m\zeta}\right)  ^{2n+|k|+N}}\nonumber\\
& \times{_{1}F_{1}}\left(  2n+|k|+N;N;\frac{NKK_{m\zeta}}{\frac{\Omega
}{2\sigma_{n}^{2}}+K_{m\zeta}}\right)  ,\label{cdf_C}%
\end{align}
where $K_{m\zeta}=\frac{1+K}{E_{m}+\zeta}$.It can be noticed that $\left(
\frac{E_{m}}{\zeta}\right)  ^{n}$ in this CDF expression is converged only
with $\frac{E_{m}}{\zeta}<1$ when $n$ goes to infinity. Therefore, it will be
used to calculate the error probability for $\Pr\left(  \zeta\geq\frac
{E_{m}+E_{m+1}}{2}\right)  $.


Similar to the case for $\zeta<E_{m}$, $B$ and $C$ for $\zeta>E_{m}$ can be
asymptotically represented as
\begin{equation}
B^{\infty}=1-{\binom{2N-1}{N}}\left(  \frac{K_{\zeta}e^{-K}}{\frac{\Omega
}{2\sigma_{n}^{2}}+K_{\zeta}}\cdot\frac{\frac{E_{m}}{\zeta}}{\left(
\frac{E_{m}}{\zeta}-1\right)  ^{2}}\right)  ^{N}%
\end{equation}
and%
\begin{align}
\acute{C}^{\infty}  & =\left(  \frac{K_{m\zeta}e^{-K}}{\frac{\Omega}%
{2\sigma_{n}^{2}}+K_{m\zeta}}\right)  ^{N}\sum_{k=1-N}^{N-1}\left(
\frac{\sqrt{E_{m}\zeta}}{E_{m}+\zeta}\right)  ^{|k|}\left(  \frac{E_{m}}%
{\zeta}\right)  ^{k}\sum_{n=0}^{\infty}\frac{(2n+|k|+N-1)!}{n!(n+|k|)!(N-1)!}%
\left(  \frac{\sqrt{E_{m}\zeta}}{E_{m}+\zeta}\right)  ^{2n}\nonumber\\
& =\left(  \frac{K_{m\zeta}e^{-K}}{\frac{\Omega}{2\sigma_{n}^{2}}+K_{m\zeta}%
}\right)  ^{N}\sum_{k=1-N}^{N-1}{\binom{N+|k|-1}{N-1}}\left(  \frac
{\sqrt{E_{m}\zeta}}{E_{m}+\zeta}\right)  ^{|k|}\left(  \frac{E_{m}}{\zeta
}\right)  ^{k}\nonumber\\
& \times{_{2}F_{1}}\left(  \frac{|k|+N}{2},\frac{|k|+N+1}{2},|k|+1,\frac
{4E_{m}\zeta}{(E_{m}+\zeta)^{2}}\right)  .\label{asymC}%
\end{align}

\subsection{The CDF for $E_{m}=0$ when $\zeta>E_{m}$}

As the CDF in (\ref{E-CDF-Kihong-00}) is not valid for $E_{0}$, the CDF
$F_{\zeta}(\zeta|E_{0})$ is derived separately. Towards this end, the
conditional CDF given $x$ can be rewritten as
\begin{align}
F_{\zeta}\left(  \zeta|x,E_{0}\right)   & =1-Q_{N}\left(  0,\sqrt{\frac{\zeta
x}{\sigma_{n}^{2}}}\right) \nonumber\\
& =1-\frac{\Gamma\left(  N,\frac{\zeta x}{2\sigma_{n}^{2}}\right)  }%
{\Gamma(N)}\nonumber\\
& =1-\exp\left(  -\frac{\zeta x}{2\sigma_{n}^{2}}\right)  \sum_{k=0}%
^{N-1}\frac{\left(  \frac{\zeta x}{2\sigma_{n}^{2}}\right)  ^{k}}{k!}.
\end{align}
The unconditional CDF is computed as
\begin{align}
F_{\zeta}(\zeta|E_{0})  & =1-\sum_{k=0}^{N-1}\frac{\left(  \frac{\zeta
}{2\sigma_{n}^{2}}\right)  ^{k}}{k!}\int_{0}^{\infty}x^{k}\exp\left(
-\frac{\zeta x}{2\sigma_{n}^{2}}\right)  f(x)dx\nonumber\\
& =1-\sum_{k=0}^{N-1}{\binom{N+k-1}{k}}\left(  K_{\zeta}e^{-K}\right)
^{N}\frac{\left(  \frac{\Omega}{2\sigma_{n}^{2}}\right)  ^{k}}{\left(
\frac{\Omega}{2\sigma_{n}^{2}}+K_{\zeta}\right)  ^{k+N}}.
\end{align}
Asymptotically, this CDF can be represented as
\[
F_{\zeta}^{\infty}(\zeta|E_{0})=1-{\binom{2N-1}{N}}\left(  \frac{K_{\zeta
}e^{-K}}{\frac{\Omega}{2\sigma_{n}^{2}}+K_{\zeta}}\right)  ^{N}.
\]
By substituting the CDFs into average SER, a closed form can be obtained form
by summation of infinite series. Moreover, by replacing the obtained CDFs with
the asymptotic CDFs, the asymptotic average SER can be obtained in a closed-form.

\section{Numerical Results}

This section presents analytical and simulation results of MASK modulation
with coherent, noncoherent and amplitude-coherent detection in flat Rician
fading channels. Moreover, the AC detection is evaluated using the optimum,
suboptimum, and the heuristic detectors with single and antennas reception
using various modulation orders. The Monte Carlo simulation results are
obtained by generating ${10}^{7}$ realizations and the average SNR is defined
as $SNR=\frac{\Omega\text{ }\bar{P}_{s}}{2\sigma_{h}^{2}}$, where $\bar{E}%
_{s}=\frac{1}{M}\sum_{m=0}^{M-1}E_{m}$ is the average transmission power. In
the analytical results, the summations with infinite limits are truncated
where 20 terms are used. For the results included in this section $\bar{P}%
_{s}$ and $\Omega$ are normalized to $1$. The figures' legends are using the
following abbreviations, simulation (Sim.), analytical (Anal.), coherent
detector (Coh.), noncoherent detector (NC), near-optimum AC (AC-NO),
suboptimum AC (AC-SO), AC heuristic (AC-H), and asymptotic (Asymp.).

\begin{figure}[ptb]%
\centering
\includegraphics[
height=3.4411in,
width=3.4411in
]%
{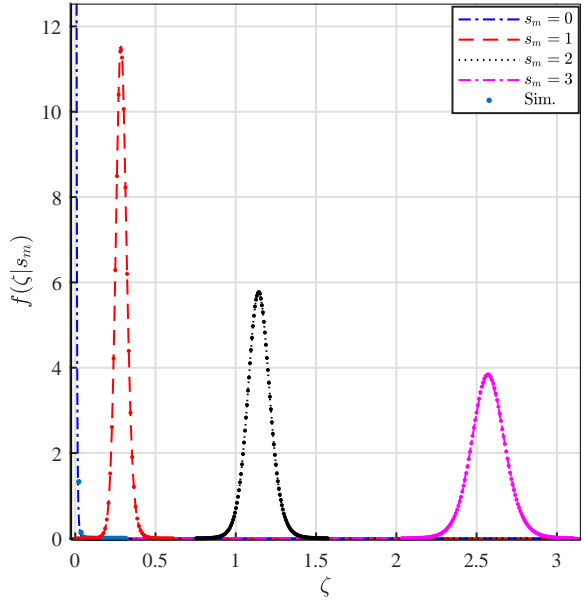}%
\caption{Analytical and simulated conditional PDF $f\left(  \zeta
|E_{m}\right)  $ for $M=4$, $K=10$ and $SNR=27$ dB.}%
\label{F-PDF-Zeta}%
\end{figure}

Fig. \ref{F-PDF-Zeta} shows the analytical and simulated conditional PDF
$f\left(  \zeta|E_{m}\right)  $, for the case of $M=4$, $K=10$ and $SNR=27$
$dB $. As can be noted from the figure, the overlap between the conditional
PDF for different $E_{m}$ values at high SNRs is negligible, and thus, the
transmitted signal can be recovered reliably by using the suitable threshold
as described in (\ref{E-Thresh_Heu}) for the heuristic AC detector. However,
because the conditional PDFs are not identical and not equally spaced, the
probability of error given $E_{m}$ will not be equal. Consequently, the
amplitudes of the transmitted signals can be optimized to minimize the BER.
Nevertheless, the improvement that would be gained is generally limited as
reported for the Rayleigh fading case \cite{Bariah-TVT}.%

\begin{figure}[ptb]%
\centering
\includegraphics[
height=3.4411in,
width=3.4411in
]%
{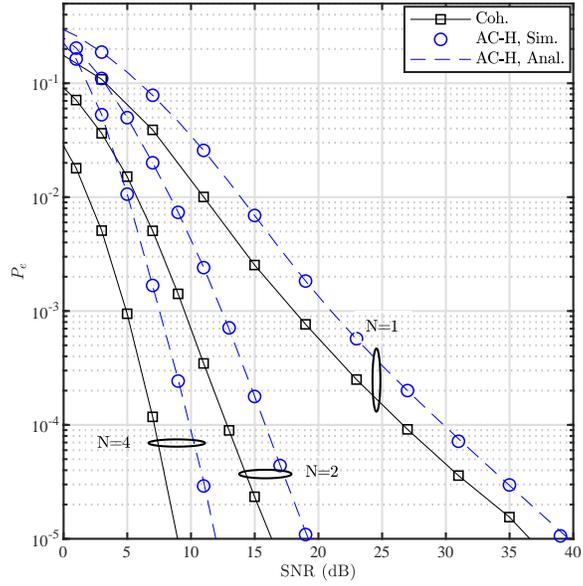}%
\caption{Analytical and simulated SER of the AC heuristic (AC-H) and coherent
detectors using $N=1$, $2$, $4$, $M=2$, and $K=4$.}%
\label{F-Diff-N-Heur-Coh-OOK}%
\end{figure}

\begin{figure}[ptb]%
\centering
\includegraphics[
height=3.4411in,
width=3.4411in
]%
{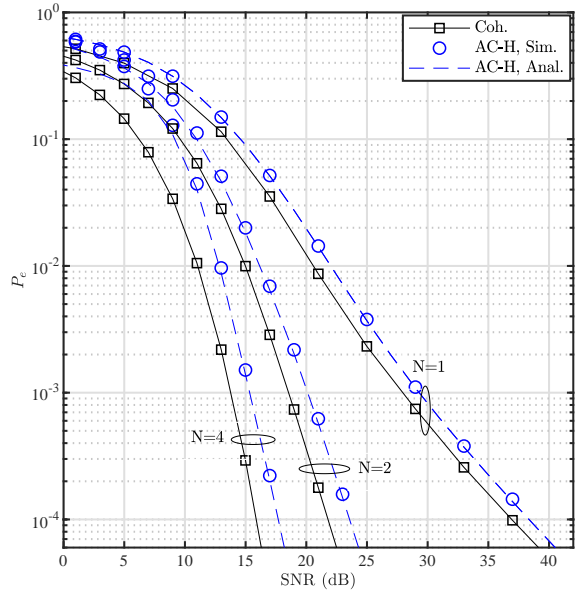}%
\caption{Analytical and simulated SER of the AC heuristic (AC-H) and coherent
detectors using $N=1$, $2$, $4$, $M=4$, and $K=4$.}%
\label{F-Diff-N-Heur-Coh-4ASK}%
\end{figure}

Figs. \ref{F-Diff-N-Heur-Coh-OOK} and \ref{F-Diff-N-Heur-Coh-4ASK} compare the
SER of the coherent and heuristic AC detectors using $M=2$ and $4$,
respectively. The figures also show the SER when the number of receiving
antennas $N=1$, $2$, and $4.$The Rician factor for both figures is fixed at
$K=4$. The results presented in both figures show that the simulated SER
perfectly matches the analytical SER for all the considered $M$ and $N$
values. Comparing the coherent and AC heuristic detector for the case of $M=2
$ in Fig. \ref{F-Diff-N-Heur-Coh-OOK} show that the coherent detector
outperforms the AC detector by about 3 dB at $P_{e}=5\times10^{-5}$. For the
case of $M=4$ shown in Fig. \ref{F-Diff-N-Heur-Coh-4ASK}, the difference
between the coherent and AC Heuristic becomes smaller and dependent on $N$.
More specifically, the difference becomes about $1.4$, $1.7$ and $2.0$ dB for
$N=1$, $2$, and $3$, respectively, at $P_{e}=5\times10^{-5}$.%


\begin{figure}[ptb]%
\centering
\includegraphics[
height=3.4411in,
width=3.4411in
]%
{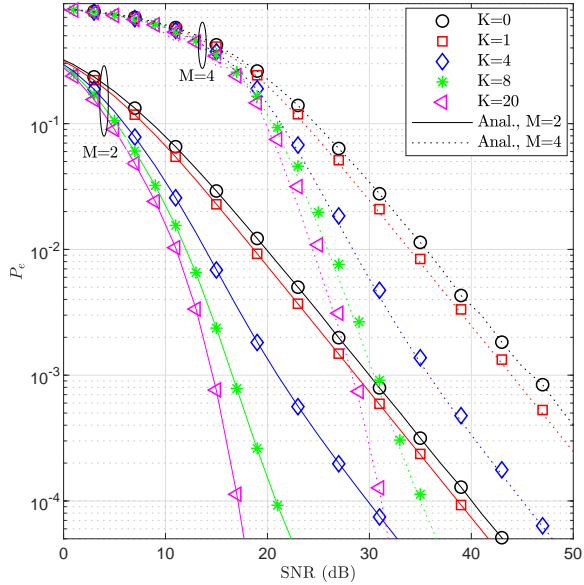}%
\caption{Analytical and simulated SER of the AC heuristic (AC-H) detector
using $M=2$ and $4$ for different values of $K$, $N=1$.}%
\label{F_diff_k_n1_heur_m2_m4}%
\end{figure}

Fig. \ref{F_diff_k_n1_heur_m2_m4} illustrates the impact of the Rician factor
$K$ on the SER for the cases of $M=2$ and $4$, for a single receiving antenna,
$N=1$. The results for $K=0$ are considered as the worst case scenario where
the channel becomes Rayleigh. As can be depicted from the figure, the SER of
the AC detector may improve substantially for large values of $K$.
Nevertheless, the SER improvement gained by increasing $K$ is higher for $M=2$
as compared with the $M=4$ case, which is due to the fact that higher order
modulations are more sensitive to AWGN, and thus, the fading will be less
dominant as compared to low order modulations. For example, the SER
improvement by increasing $K$ from $1$ to $20$ is about $20$ dB for $M=2$,
while it is about $18$ dB for $M=4$, at $P_{e}=4\times10^{-4}$.%

\begin{figure}[ptb]%
\centering
\includegraphics[
height=3.4411in,
width=3.4411in
]%
{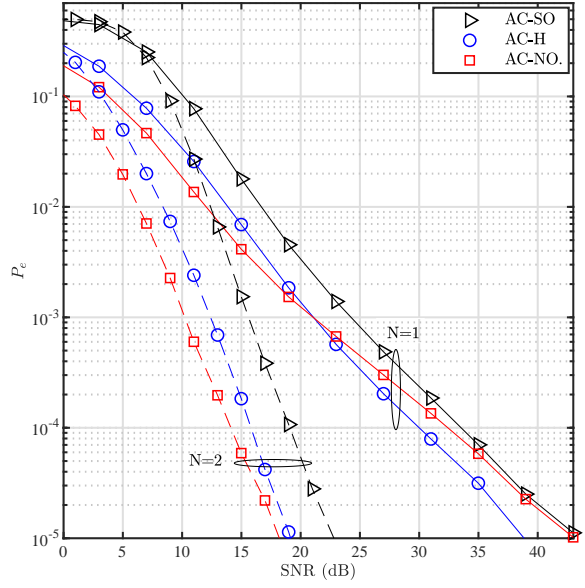}%
\caption{SER for of the near optimum (AC-NO), suboptimum (AC-SO), and
heuristic AC (AC-H) detectors using $N=1$, $2$, $M=2$, and $K=4$.}%
\label{F_her_subopt_opt_acd_diff_n_m2}%
\end{figure}

\begin{figure}[ptb]%
\centering
\includegraphics[
height=3.4411in,
width=3.4411in
]%
{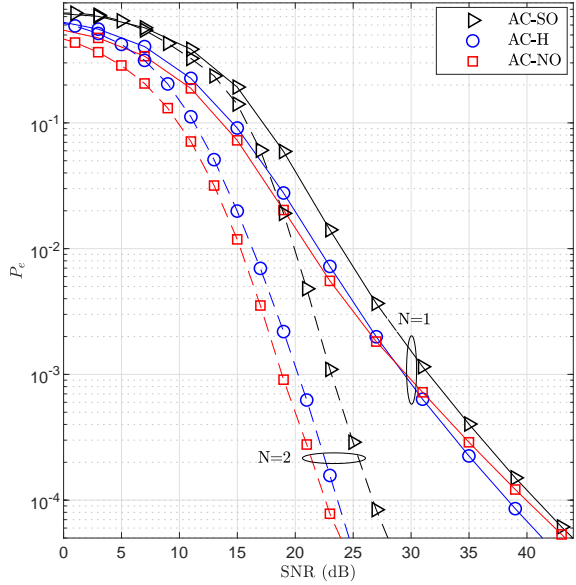}%
\caption{SER for of the near optimum (AC-NO), suboptimum (AC-SO) and heuristic
AC (AC-H) detectors using $N=1$, $2 $, $M=4$, and $K=4$.}%
\label{fig6_her_subopt_opt_acd_diff_n_m4}%
\end{figure}

Figs. \ref{F_her_subopt_opt_acd_diff_n_m2} and
\ref{fig6_her_subopt_opt_acd_diff_n_m4}\ compare the SER performance of the
near optimum, suboptimum and heuristic AC detectors for $M=2$ and $4$,
respectively, and the results are obtained using $K=4$. As shown in both
figures, the heuristic detector outperforms the suboptimum for all cases. For
$M=2$, the difference is about $3.5$ and $3.7$ dB for $N=1$ and $2$,
respectively. For $M=4$, the difference is about $2.5$ and $3$ dB for $N=1$
and $2$, respectively. As expected, the near-optimum detector outperforms the
heuristic for $SNR\lesssim21$ dB for $M=2$ and $N=1$, which corresponds the
low and moderate SNRs. For $N=2$, the system SNR is generally much smaller
that the $N=1$ case, and hence, near-optimum detector outperforms the
heuristic. The SER for the $M=4$ case in Fig.
\ref{fig6_her_subopt_opt_acd_diff_n_m4} is generally similar to the $M=2$
case, except that the cross-over point is shifted to $SNR\approx29$ dB.
Consequently, the heuristic detector offers the best compromise between SER
and computational complexity as compared to the suboptimum and near-optimum
detectors.%

\begin{figure}[ptb]%
\centering
\includegraphics[
height=3.4411in,
width=3.4411in
]%
{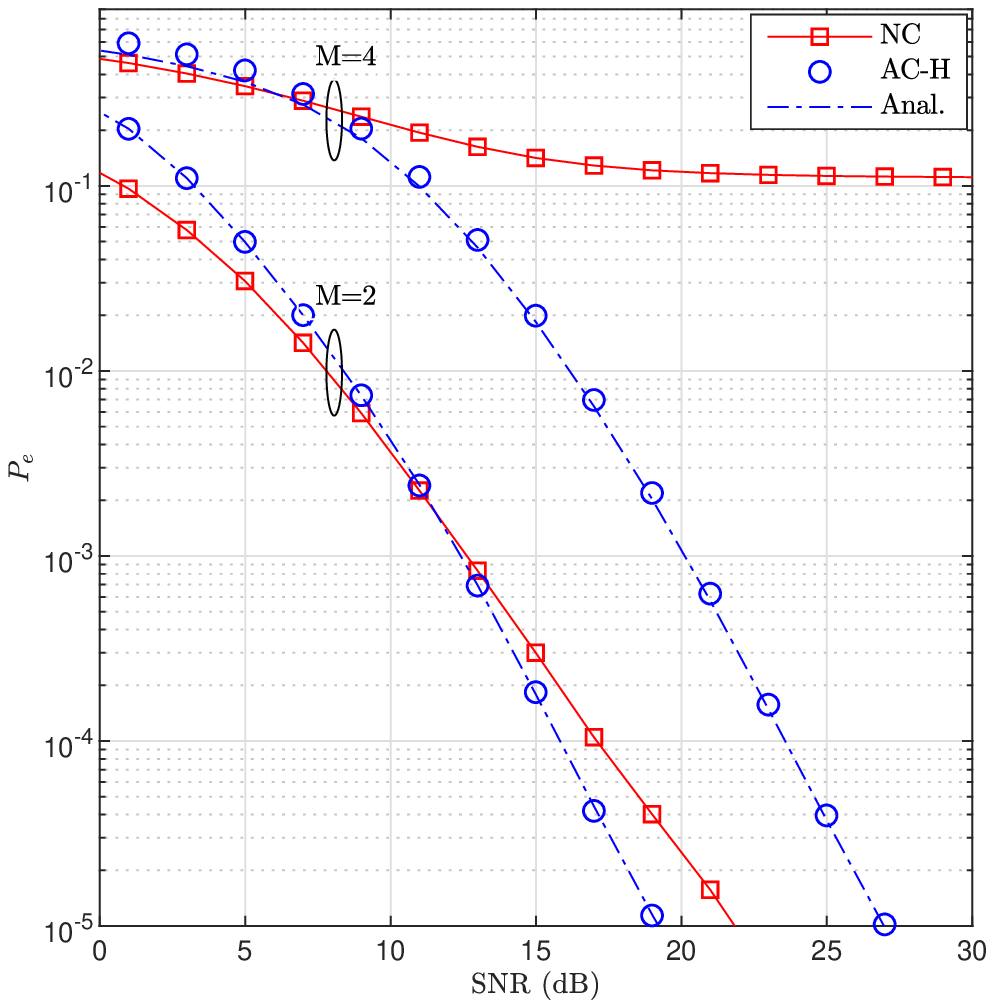}%
\caption{The SER for the noncoherent (NC) and heuristic (AC-H) detectors using
$M=1$ and $2$, $N=1$ and $K=4$.}%
\label{fig7_ncd_heur_n2_m2_m4}%
\end{figure}

Fig. \ref{fig7_ncd_heur_n2_m2_m4} presents the system SER using the optimum
noncoherent and the heuristic AC detector given that $M=2$ and $4$, $N=1$, and
$K=4$. As can be noted from the figure, the noncoherent detector outperforms
the heuristic for SNRs less than $11$ and $5$ dB, for $M=2$ and $4$
respectively. Nevertheless, the noncoherent detector SER deteriorates severely
for $M>2$ where an error floor is observed at $SER\sim10^{-1}$. Moreover, it
is worth noting that the optimum noncoherent detector requires prior knowledge
of the channel statistical values.%

\begin{figure}[ptb]%
\centering
\includegraphics[
height=3.4411in,
width=3.4411in
]%
{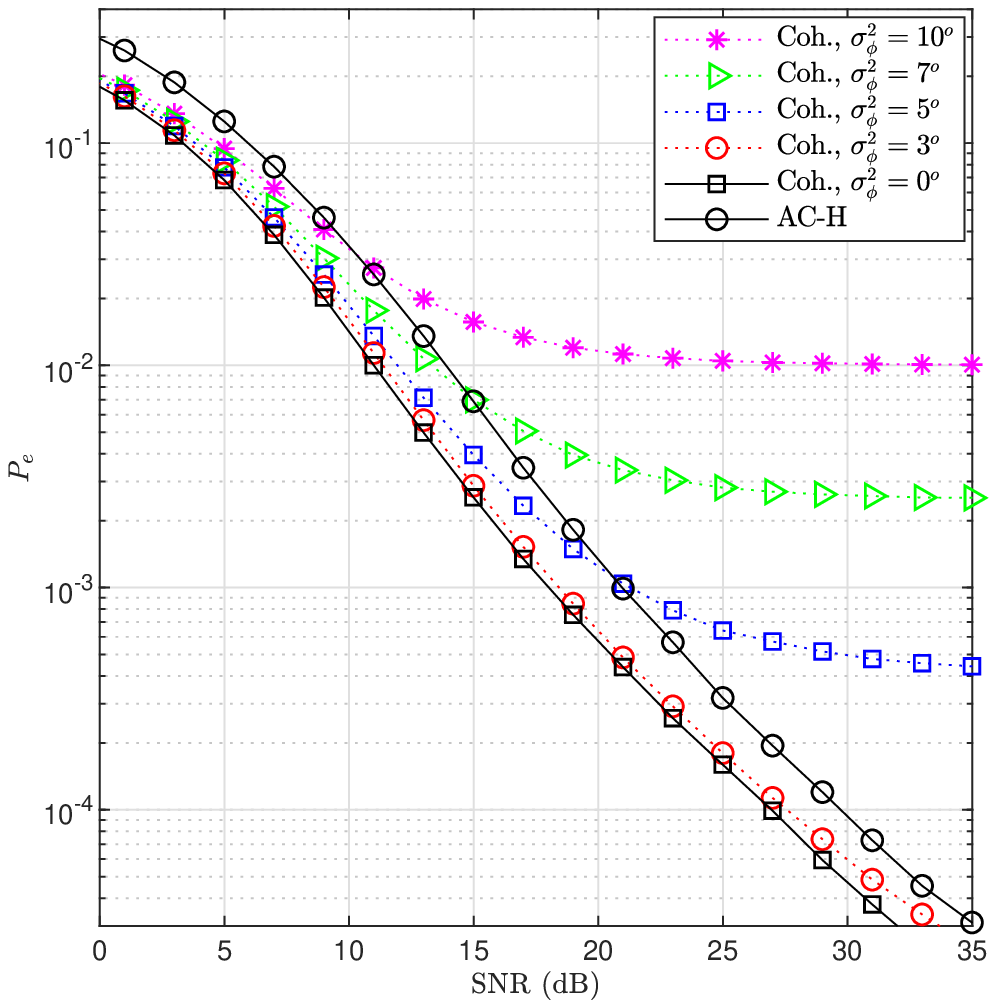}%
\caption{The effect of the phase noise on the heuristic AC (AC-H) and coherent
detectors for $N=1$, where $M=2$, $K=4$.}%
\label{fig8_phase_noise_m2}%
\end{figure}

\begin{figure}[ptb]%
\centering
\includegraphics[
height=3.4411in,
width=3.4411in
]%
{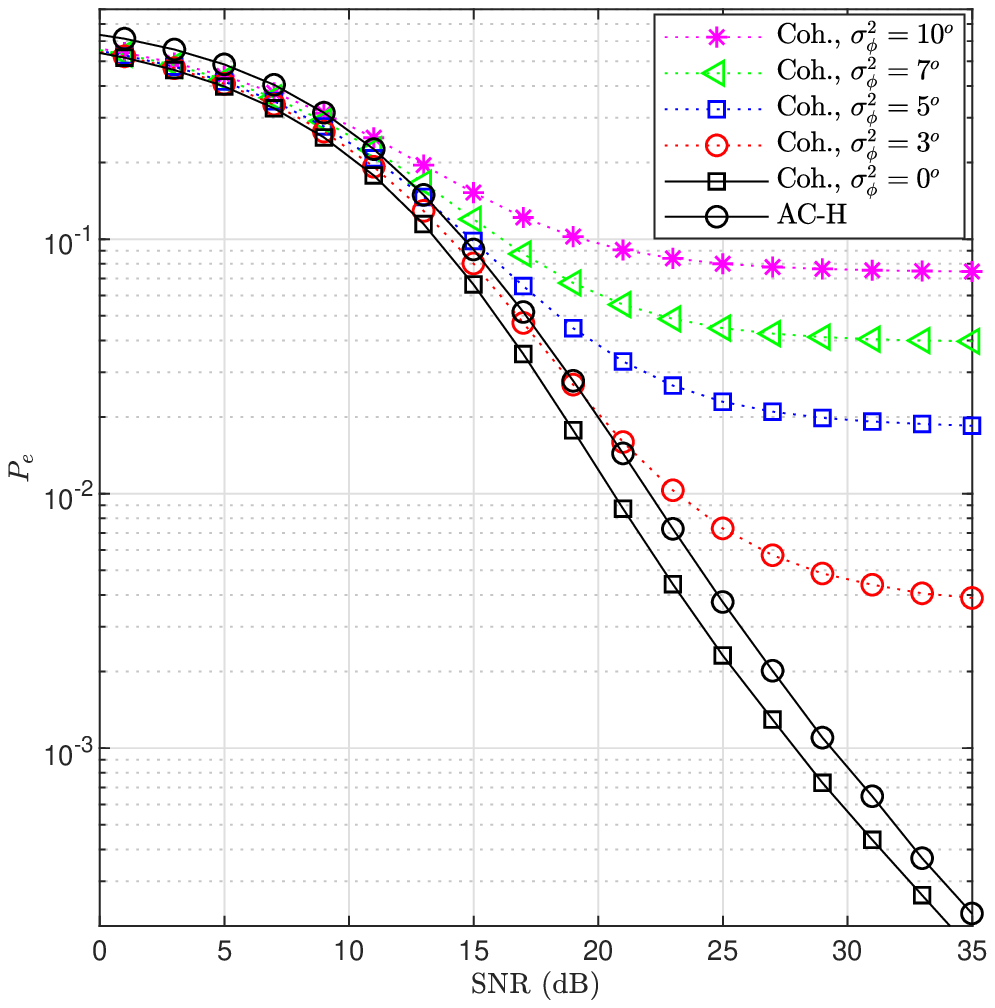}%
\caption{The effect of the phase noise on the heuristic AC (AC-H) and coherent
detectors for $N=1$, where $M=4$, $K=4$.}%
\label{fig9_phase_noise_m4}%
\end{figure}

Figs. \ref{fig8_phase_noise_m2} and \ref{fig9_phase_noise_m4} show the effect
of the phase noise on both the heuristic and coherent detectors for the cases
of $M=2$ and $4$, respectively. The parameters for the two figures are $N=1$
and $K=4$ while the phase noise is modeled as a Tikhonov random variable with
variance $\sigma_{\phi}^{2}=\left[  0\text{, }3\text{, }5\text{, }7\text{,
}10\right]  $. In both figures, the SER of the AC detector is represented by a
single curve because it is immune to phase noise. The results in Figs.
\ref{fig8_phase_noise_m2} and \ref{fig9_phase_noise_m4} show clearly the
advantage of the heuristic detector in the presence of phase noise,
particularly at high SNRs, where coherent detector exhibits SER error floors.
As expected, the $M=4$ case is more sensitive than the $M=2$, even for very
small values of $\sigma_{\phi}^{2}$.%

\begin{figure}[ptb]%
\centering
\includegraphics[
height=3.4411in,
width=3.4411in
]%
{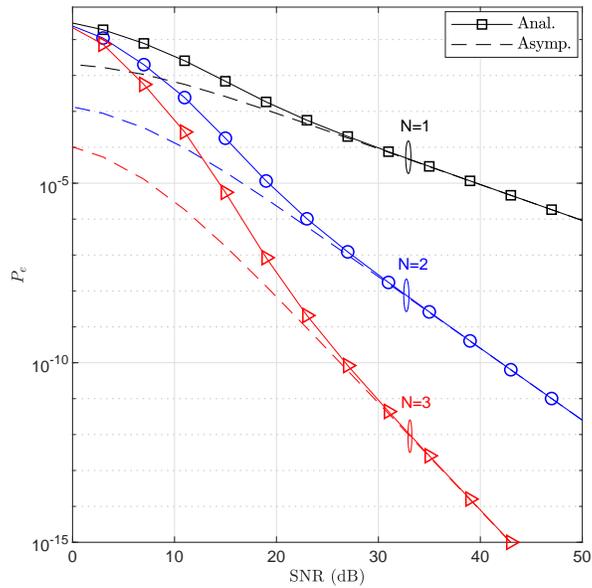}%
\caption{Asymptotic SER using $M=2$, $K=4$, and $N=1,2$ and $3$.}%
\label{F-Asympt}%
\end{figure}

Fig. \ref{F-Asympt} compares the asymptotic and analytical SERs using $M=2$,
$K=4,$ $N=1,$ $2$ and $3$. As can be noted from the figure, the asymptotic SER
provides accurate results for\ the SER at high SNRs, and thus, it can be used
to simplify the SER analysis.%

\section{Conclusion and Future Work}

In this paper, the SER performance of MASK modulation with amplitude-coherent
detection has been considered over flat Rician fading channels with receiver
diversity. The optimum, near-optimum and suboptimum amplitude-coherent
detectors were derived for the Rician channel and their SER was compared with
the coherent and noncoherent detectors for various modulation orders and
number of receiving antennas. The SER of the heuristic detector was derived
analytically using different approaches, and the asymptotic SER was derived
for high SNRs. The obtained analytical and simulation results confirm that the
amplitude-coherent detection offers the SER performance that is comparable to
the coherent detection, but without the need for a prior knowledge of the
channel phase. Consequently, the computational complexity of the amplitude
coherent detector is much less than the coherent detector and it is more
robust to phase noise and phase estimation errors.

\section*{Appendix I: Evaluating the integral $\mathcal{I}_{\theta}$}

The integral $\mathcal{I}_{\theta}$\ given in (\ref{I-theta}) can be written
as%
\begin{equation}
\mathcal{I}_{\theta}=\frac{1}{2\pi}\int_{-\pi}^{\pi}\exp\left(  2\sqrt
{K\left(  K+1\right)  }\cos\left(  \theta-\phi\right)  +\frac{\alpha s_{m}%
}{\sigma_{n}^{2}}\left\vert r\right\vert \cos\left(  \theta-\theta_{\text{r}%
}\right)  \right)  d\theta.
\end{equation}
Applying the trigonometric identity $\cos\left(  \theta-\phi\right)
=\cos\theta\cos\phi+\sin\theta\sin\phi$ yields%
\begin{align}
\mathcal{I}_{\theta}  & =\frac{1}{2\pi}\int_{-\pi}^{\pi}\exp\left(
2\sqrt{K\left(  K+1\right)  }\left(  \cos\left(  \theta\right)  \cos\left(
\phi\right)  +\sin\left(  \theta\right)  \sin\left(  \phi\right)  \right)
+\frac{\alpha s_{m}}{\sigma_{n}^{2}}\left\vert r\right\vert \left(
\cos\left(  \theta\right)  \cos\left(  \theta_{\text{r}}\right)  +\sin\left(
\theta\right)  \sin\left(  \theta_{\text{r}}\right)  \right)  \right)
d\theta\nonumber\\
& =\frac{1}{2\pi}\int_{-\pi}^{\pi}\underset{g\left(  \theta\right)
}{\underbrace{\exp\left(  \left(  A+B\alpha s_{m}\left\vert r\right\vert
\right)  \cos\left(  \theta\right)  +\left(  C+D\alpha s_{m}\left\vert
r\right\vert \right)  \sin\left(  \theta\right)  \right)  }}d\theta
\label{I_theta}%
\end{align}
where $A=2\sqrt{K\left(  K+1\right)  }\cos\left(  \phi\right)  $, $B=\frac
{1}{\sigma_{n}^{2}}\cos\left(  \theta_{\text{r}}\right)  $, $C=2\sqrt{K\left(
K+1\right)  }\sin\left(  \phi\right)  $ and $D=\frac{1}{\sigma_{n}^{2}}%
\sin\left(  \theta_{\text{r}}\right)  $.

It should be observed that the function $g\left(  \theta\right)  $ is not
symmetric at $\theta=0$, and thus $\int_{-\pi}^{\pi}g\left(  \theta\right)
d\theta\neq2\int_{0}^{\pi}g\left(  \theta\right)  d\theta$. However, the main
interest is to evaluate the integral of $g\left(  \theta\right)  $\ rather
than calculating the value of $g\left(  \theta\right)  $ itself. Therefore, a
mathematical manipulation can be made to force symmetry of $g\left(
\theta\right)  $\ at $\theta=0$ without affecting the value of the integral
$\mathcal{I}_{\theta}$. Towards this goal, first the value of $\theta$ at
which the function $g\left(  \theta\right)  $\ has a global maximum, $\varphi
$, is evaluated, and then the function is shifted by the same value in order
to have the maximum at $\theta=0$. Then, the integral can be divided into two
intervals
\begin{equation}
\int_{-\pi}^{\pi}g\left(  \theta\right)  d\theta=\int_{-\pi}^{\pi}g\left(
\theta+\varphi\right)  d\theta=2\int_{0}^{\pi}g\left(  \theta+\varphi\right)
d\theta.
\end{equation}
Then, the resulting integral is evaluated numerically by applying
Gauss-Chebyshev quadrature rule.%
\begin{equation}
\frac{dg\left(  \theta\right)  }{d\theta}=\left[  -\left(  A+B\alpha
s_{m}\left\vert r\right\vert \right)  \sin\theta+\left(  C+D\alpha
s_{m}\left\vert r\right\vert \right)  \cos\theta\right]  \exp\left(  \left(
A+B\alpha s_{m}\left\vert r\right\vert \right)  \cos\theta+\left(  C+D\alpha
s_{m}\left\vert r\right\vert \right)  \sin\theta\right)  .
\end{equation}
To find $\varphi\in(-\pi,\pi)$, the unique root of $\frac{dg\left(
\theta\right)  }{d\theta}=0$ is calculated,%
\begin{equation}
\varphi=\tan^{-1}\left(  \frac{C+D\alpha s_{m}\left\vert r\right\vert
}{A+B\alpha s_{m}\left\vert r\right\vert }\right)  \text{.}%
\end{equation}
Thus, the symmetric function around $\theta=0$, $g\left(  \theta
+\varphi\right)  $, can be written as%
\begin{equation}
g\left(  \theta+\varphi\right)  =\exp\left(  \left(  A+B\alpha s_{m}\left\vert
r\right\vert \right)  \cos\left(  \theta+\varphi\right)  +\left(  C+D\alpha
s_{m}\left\vert r\right\vert \right)  \sin\left(  \theta+\varphi\right)
\right)  .
\end{equation}
Therefore, the integral $\mathcal{I}_{\theta}$ given in (\ref{I_theta}) can be
rewritten as as%
\begin{equation}
\mathcal{I}_{\theta}=2\int_{0}^{\pi}g\left(  \theta+\varphi\right)  d\theta.
\end{equation}
Evaluating the integral $\mathcal{I}_{\theta}$\ by substitution, setting
$y=\cos\left(  \theta\right)  $, yields%
\begin{multline}
\mathcal{I}_{\theta}=\frac{1}{\pi}\int_{-1}^{1}\frac{1}{\sqrt{1-y^{2}}}%
\exp\left\{  \left(  A+B\alpha s_{m}\left\vert r\right\vert \right)
\cos\left(  \varphi+\cos^{-1}\left(  y\right)  \right)  \right. \nonumber\\
\left.  +\left(  C+D\alpha s_{m}\left\vert r\right\vert \right)  \sin\left(
\varphi+\cos^{-1}\left(  y\right)  \right)  \right\}  dy
\end{multline}
which can be efficiently solved applying Gauss-Chebyshev quadrature rules
\cite{Stegun}%
\begin{equation}
\mathcal{I}_{\theta}=\frac{1}{L}\sum\limits_{l=1}^{n}\exp\left[  \left(
A+B\alpha s_{m}\left\vert r\right\vert \right)  f_{\text{c}}\left(
y_{l}\right)  +\left(  C+D\alpha s_{m}\left\vert r\right\vert \right)
f_{\text{s}}\left(  y_{l}\right)  \right]  ,
\end{equation}
where $f_{\text{c}}\left(  y_{l}\right)  =\cos\left(  \varphi+\cos^{-1}\left(
y_{l}\right)  \right)  =\cos\left(  \varphi+\frac{2l-1}{2L}\pi\right)  $,
$f_{\text{s}}\left(  y_{l}\right)  =\sin\left(  \varphi+\cos^{-1}\left(
y_{l}\right)  \right)  =\sin\left(  \varphi+\frac{2l-1}{2L}\pi\right)  $ and
$y_{l}=\cos\left(  \frac{2l-1}{2L}\pi\right)  $.

\section*{Appendix II: Evaluating $F_{\zeta}\left(  \zeta|E_{m}\right)  $,
$m>0$}

To obtain the PDF $\textup{f}\left(  \zeta|E_{m}\right)  $, averaging
$\textup{f}\left(  \zeta|x,E_{m}\right)  $ over $\textup{f}\left(  x\right)  $
is needed, which yields%
\begin{align}
f\left(  \zeta|E_{m}\right)   & =\int_{0}^{\infty}f\left(  \zeta
|x,E_{m}\right)  \text{ }f\left(  x\right)  dx\\
& =\frac{\exp\left(  -\frac{\lambda}{2}\right)  }{4\sigma_{n}^{2}\sigma
_{h}^{2}}E_{m}^{-0.5\left(  N-1\right)  }\left(  \sigma_{h}^{2}\lambda\right)
^{-0.5\left(  N-1\right)  }\zeta^{0.5\left(  N-1\right)  }\nonumber\\
& \times\int_{0}^{\infty}x^{0.5\left(  N+1\right)  }\exp\left(  -\left(
\frac{\zeta+E_{m}}{2\sigma_{n}^{2}}+\frac{1}{2\sigma_{h}^{2}}\right)
x\right)  \textup{I}_{N-1}\left(  \frac{\sqrt{E_{m}\text{ }\zeta}}{\sigma
_{n}^{2}}x\right)  \textup{I}_{N-1}\left(  \sqrt{\frac{\lambda}{\sigma_{h}%
^{2}}}\sqrt{x}\right)  dx.\label{f_zeta_sum1}%
\end{align}
Substituting the two approximations (\ref{apprx1}) and (\ref{apprx2}) in
(\ref{f_zeta_sum1}) yields%
\[
f\left(  \zeta|E_{m}\right)  =\frac{\lambda^{-0.5\left(  N-1\right)  }%
\exp\left(  -\frac{\lambda}{2}\right)  }{4\sigma_{n}^{2}\sqrt{\frac{2\pi
}{\sigma_{n}^{2}}}}E_{m}^{-0.5\left(  N-1\right)  }\frac{\zeta^{0.5\left(
N-1\right)  }\left(  I_{A}+I_{B}\right)  }{\sqrt[4]{E_{m}\text{ }\zeta}},
\]
where $I_{A}$ and $I_{B}$ are given by%
\begin{align}
I_{A}  & =\int_{0}^{\infty}x^{0.5N}\exp\left(  -C_{\zeta}x\right)  \sum
_{l=0}^{\infty}\frac{\left(  \frac{\lambda}{4\sigma_{h}^{2}}\right)
^{\frac{2l+N-1}{2}}}{l!\Gamma\left(  l+N\right)  }x^{\frac{2l+N-1}{2}}dx\\
I_{B}  & =\int_{0}^{\infty}x^{0.5N}\exp\left(  -C_{\zeta}x\right)  \sum
_{l=0}^{\infty}\frac{\left(  \frac{\lambda}{4\sigma_{h}^{2}}\right)
^{\frac{2l+N-1}{2}}x^{\frac{2l+N-1}{2}}}{l!\Gamma\left(  l+N\right)  }%
\sum_{q=1}^{Q}\frac{\left(  -1\right)  ^{q}}{x^{q}}\frac{\prod_{k=1}%
^{q}\left[  4\left(  N-1\right)  ^{2}-\left(  2k-1\right)  ^{2}\right]
}{q!8^{q}\left(  \frac{\sqrt{E_{m}\text{ }\zeta}}{\sigma_{n}^{2}}\right)
^{q}}dx.
\end{align}
The integral $I_{A}$ can be evaluated as%

\begin{align}
I_{A}  & =\int_{0}^{\infty}x^{0.5N}\exp\left(  -C_{\zeta}x\right)  \sum
_{l=0}^{\infty}\frac{1}{l!\Gamma\left(  l+N\right)  }\left(  \frac{\lambda
}{4\sigma_{h}^{2}}\right)  ^{\frac{2l+N-1}{2}}x^{\frac{2l+N-1}{2}%
}dx\nonumber\\
& =\sum_{l=0}^{\infty}\frac{1}{l!\Gamma\left(  l+N\right)  }\left(
\frac{\lambda}{4\sigma_{h}^{2}}\right)  ^{\frac{2l+N-1}{2}}\int_{0}^{\infty
}x^{l+N-0.5}\exp\left(  -C_{\zeta}x\right)  dx.\label{E-I-A}%
\end{align}
By substituting $y=\left(  \frac{\zeta+E_{m}}{2\sigma_{n}^{2}}+\frac
{1}{2\sigma_{h}^{2}}-\frac{\sqrt{E_{m}\text{ }\zeta}}{\sigma_{n}^{2}}\right)
x $ in (\ref{E-I-A}), \ can be evaluated as%
\begin{align}
I_{A}  & =\sum_{l=0}^{\infty}\frac{1}{l!\Gamma\left(  l+N\right)  }\left(
\frac{\lambda}{4\sigma_{h}^{2}}\right)  ^{\frac{2l+N-1}{2}}\frac{1}{C_{\zeta
}^{l+N+0.5}}\int_{0}^{\infty}y^{l+N-0.5}\exp\left(  -y\right)  dy\nonumber\\
& =\sum_{l=0}^{\infty}\frac{1}{l!\Gamma\left(  l+N\right)  }\left(
\frac{\lambda}{4\sigma_{h}^{2}}\right)  ^{\frac{2l+N-1}{2}}\frac{1}{C_{\zeta
}^{l+N+0.5}}\Gamma\left(  l+N+0.5\right) \nonumber\\
& =\sum_{l=0}^{\infty}\frac{A_{l}}{C_{\zeta}^{l+N+0.5}},
\end{align}

The integral $I_{B}$ can be evaluated as following%
\begin{align}
I_{B}  & =\int_{0}^{\infty}x^{0.5N}\exp\left(  -C_{\zeta}x\right)  \sum
_{q=1}^{Q}\frac{\left(  -1\right)  ^{q}}{x^{q}}\frac{\prod_{k=1}^{q}\left[
4\left(  N-1\right)  ^{2}-\left(  2k-1\right)  ^{2}\right]  }{q!8^{q}\left(
\frac{\sqrt{E_{m}\text{ }\zeta}}{\sigma_{n}^{2}}\right)  ^{q}}\nonumber\\
& \times\sum_{l=0}^{\infty}\frac{1}{l!\Gamma\left(  l+N\right)  }\left(
\frac{\lambda}{4\sigma_{h}^{2}}\right)  ^{\frac{2l+N-1}{2}}x^{\frac{2l+N-1}%
{2}}dx\\
& =\sum_{q=1}^{Q}\sum_{l=0}^{L=\infty}\frac{\left(  -1\right)  ^{q}}%
{l!\Gamma\left(  l+N\right)  }\left(  \frac{\lambda}{4\sigma_{h}^{2}}\right)
^{\frac{2l+N-1}{2}}\frac{\prod_{k=1}^{q}\left[  4\left(  N-1\right)
^{2}-\left(  2k-1\right)  ^{2}\right]  }{q!8^{q}\left(  \frac{\sqrt
{E_{m}\text{ }\zeta}}{\sigma_{n}^{2}}\right)  ^{q}}\nonumber\\
& \times\int_{0}^{\infty}x^{N-q+l-0.5}\exp\left(  -C_{\zeta}x\right)  dx.
\end{align}
Substituting $y=\left(  \frac{\zeta+E_{m}}{2\sigma_{n}^{2}}+\frac{1}%
{2\sigma_{h}^{2}}-\frac{\sqrt{E_{m}\text{ }\zeta}}{\sigma_{n}^{2}}\right)  x
$\ yields%
\begin{align}
I_{B}  & =\sum_{q=1}^{Q}\sum_{l=0}^{L=\infty}\frac{\left(  -1\right)  ^{q}%
}{l!\Gamma\left(  l+N\right)  }\left(  \frac{\lambda}{4\sigma_{h}^{2}}\right)
^{\frac{2l+N-1}{2}}\left(  \frac{\prod_{k=1}^{q}\left[  4\left(  N-1\right)
^{2}-\left(  2k-1\right)  ^{2}\right]  }{q!8^{q}\left(  \frac{\sqrt
{E_{m}\text{ }\zeta}}{\sigma_{n}^{2}}\right)  ^{q}}\right) \nonumber\\
& \times\frac{1}{C_{\zeta}^{N-q+l+0.5}}\int_{0}^{\infty}y^{N-q+l-0.5}%
\exp\left(  -y\right)  dy\\
& =\sum_{q=1}^{Q}\sum_{l=0}^{\infty}\frac{B_{q}^{l}}{\zeta^{0.5q}C_{\zeta
}^{N-q+l+0.5}}.
\end{align}

Finally, $\textup{f}\left(  \zeta|E_{m}\right)  $\ is obtained as%
\begin{equation}
f\left(  \zeta|E_{m}\right)  =C\text{ }\left(  \sum_{l=0}^{\infty}\frac
{A_{l}\zeta^{0.5N-0.75}}{C_{\zeta}^{\left(  l+N+0.5\right)  }}+\sum_{q=1}%
^{Q}\sum_{l=0}^{\infty}\frac{B_{q}^{l}\zeta^{0.5N-0.5q-0.75}}{C_{\zeta
}^{\left(  N-q+l+0.5\right)  }}\right)
\end{equation}

The CDF can be evaluated as%
\begin{align}
F_{\zeta}\left(  \zeta|E_{m}>0\right)   & =\int_{0}^{\zeta}f\left(
\zeta|E_{m}\right)  d\zeta\nonumber\\
& =C\left(  \sum_{l=0}^{\infty}A_{l}\mathcal{I}_{A}^{l}+\sum_{q=1}^{Q}%
\sum_{l=0}^{\infty}B_{q}^{l}\mathcal{I}_{B}^{q,l}\right)  ,
\end{align}
where%
\begin{align}
\mathcal{I}_{A}^{l}  & =\int_{0}^{\zeta}\zeta^{0.5N-0.75}C_{\zeta}^{-\left(
l+N+0.5\right)  }d\zeta\\
\mathcal{I}_{B}^{q,l}  & =\int_{0}^{\zeta}\zeta^{0.5N-0.5q-0.75}C_{\zeta
}^{-\left(  N-q+l+0.5\right)  }d\zeta.
\end{align}
It should be observed that both integrals have the same form, and thus they
can be rewritten as%
\begin{align}
\mathcal{I}  & =\int_{0}^{\zeta}\zeta^{a}C_{\zeta}^{b}d\zeta\\
& =\int_{0}^{\zeta}\zeta^{a}\left(  \frac{\zeta}{2\sigma_{n}^{2}}-\frac
{\sqrt{E_{m}\text{ }\zeta}}{\sigma_{n}^{2}}+c\right)  ^{b}d\zeta,
\end{align}

Substituting $y=\sqrt{\zeta}$, the integrals $\mathcal{I}_{A}^{l}$\ and
$\mathcal{I}_{B}^{q,l}$\ have the following form%
\begin{equation}
\mathcal{I}=2\int_{0}^{\sqrt{\zeta}}y^{2a+1}\left(  \frac{y^{2}}{2\sigma
_{n}^{2}}-\frac{\sqrt{E_{m}}}{\sigma_{n}^{2}}y+c\right)  ^{b}dy.
\end{equation}
It should be noticed that integrals of this form can be solved in recursive
manner according to \cite[2.17, page 79]{Table}.

\section{Appendix III: Evaluating $F_{\zeta}\left(  \zeta|E_{m}\right)  $,
$m=0$}

Averaging $f\left(  \zeta|x,E_{m}\right)  $, $m=0$ over $\textup{f}\left(
x\right)  $ yields%
\begin{align}
f\left(  \zeta|E_{m}\right)   & =\int_{0}^{\infty}f\left(  \zeta
|x,E_{m}\right)  \text{ }f\left(  x\right)  dx\nonumber\\
& =\frac{\lambda_{0}^{N}\left(  \lambda_{0}\sigma_{h}^{2}\right)
^{-0.5\left(  N-1\right)  }}{2\sigma_{h}^{2}\left(  N-1\right)  !\exp\left(
\frac{\lambda_{0}}{2}\right)  }\zeta^{N-1}\int_{0}^{\infty}x^{1.5N-0.5}%
\exp\left[  -x\left(  \lambda_{0}\zeta+\frac{1}{2\sigma_{h}^{2}}\right)
\right]  \textup{I}_{N-1}\left(  \sqrt{\frac{\lambda_{0}x}{\sigma_{h}^{2}}%
}\right)  dx.\label{f_zeta_sum}%
\end{align}
Substituting the approximation given in (\ref{apprx1}) in (\ref{f_zeta_sum})
yields,%
\begin{equation}
f\left(  \zeta|E_{m}\right)  =\frac{\lambda_{0}^{N}}{\left(  N-1\right)
!}\frac{1}{2\sigma_{h}^{2}}\left(  \zeta\right)  ^{N-1}\left(  \frac
{1}{\lambda_{0}\sigma_{h}^{2}}\right)  ^{0.5\left(  N-1\right)  }\exp\left(
-\frac{\lambda_{0}}{2}\right)  \sum_{l=0}^{\infty}\frac{1}{l!\Gamma\left(
l+N\right)  }\left(  \frac{\lambda_{0}}{4\sigma_{h}^{2}}\right)
^{\frac{2l+N-1}{2}}I_{0}^{l},
\end{equation}
where%
\begin{equation}
I_{0}^{l}=\int_{0}^{\infty}x^{2N+l-1}\exp\left(  -\left(  \lambda_{0}\text{
}\zeta+\frac{1}{2\sigma_{h}^{2}}\right)  x\right)  dx.\label{E-I0-l}%
\end{equation}
Substituting $y=\left(  \lambda_{0}\text{ }\zeta+\frac{1}{2\sigma_{h}^{2}%
}\right)  x$ in (\ref{E-I0-l}), $I_{0}^{l}$ can be obtained as%
\begin{align}
I_{0}^{l}  & =\frac{1}{\left(  \lambda_{0}\text{ }\zeta+\frac{1}{2\sigma
_{h}^{2}}\right)  ^{2N+l}}\int_{0}^{\infty}\left(  y\right)  ^{2N+l-1}%
\exp\left(  -y\right)  dy\nonumber\\
& =\frac{1}{\left(  \lambda_{0}\text{ }\zeta+\frac{1}{2\sigma_{h}^{2}}\right)
^{2N+l}}\Gamma\left(  2N+l\right)  .
\end{align}
Consequently, $\textup{f}\left(  \zeta|E_{m}\right)  $\ can be expressed as
\begin{equation}
f\left(  \zeta|E_{m}\right)  =C_{0}\text{ }\zeta^{N-1}\sum_{l=0}^{\infty}%
\frac{\Gamma\left(  2N+l\right)  }{l!\Gamma\left(  l+N\right)  }\left(
\frac{\lambda_{0}}{4\sigma_{h}^{2}}\right)  ^{\frac{2l+N-1}{2}}\left(
\lambda_{0}\zeta+\frac{1}{2\sigma_{h}^{2}}\right)  ^{-\left(  2N+l\right)
}\text{, }m=0.
\end{equation}

\end{document}